\documentclass[journal]{IEEEtran}
\usepackage{graphicx}
\usepackage{float}
\usepackage{amsmath}
\usepackage{amssymb}
\usepackage{commath} % norm and absolute value
\usepackage{mathtools}
\usepackage{algorithm}
\usepackage{algpseudocode}
\usepackage{pifont}
\usepackage{float}
\usepackage{color}
\usepackage{cite}
\usepackage[export]{adjustbox}
\usepackage[nameinlink,capitalize]{cleveref}
\usepackage{parskip}
\usepackage{epstopdf}
\usepackage{url}
\usepackage[caption=false,font=footnotesize]{subfig}

\newcommand{\vect}[1]{\boldsymbol{#1}}

\begin{document}

\title{Frequency Stability Using MPC-based Inverter Power Control in Low-Inertia Power Systems}

\author{\IEEEauthorblockN{Atinuke Ademola-Idowu and Baosen Zhang}%
\thanks{The authors are with the Department of Electrical and Computer Engineering at the University of Washington, Seattle, WA 98195. Email: \{aidowu, zhangbao\}@uw.edu}%
\thanks{This work was partially supported by the National Science Foundation grant ECCS-1930605, the U.S. Department of Energy’s Office of Energy Efficiency and Renewable Energy (EERE) under Solar Energy Technologies Office (SETO) Agreement Number EE0009025 and the University of Washington Clean Energy Institute.}}
% \IEEEauthorblockA{Electrical Engineering Department,\\
% University of Washington, Seattle, WA 98195}\\
% Email: \{aidowu, zhangbao\}@uw.edu}

% \markboth{Journal of \LaTeX\ Class Files,~Vol.~14, No.~8, August~2015}%
% {Shell \MakeLowercase{\textit{et al.}}: Bare Demo of IEEEtran.cls for IEEE Journals}

\maketitle
\begin{abstract}
The electrical grid is evolving from a network consisting of mostly synchronous machines to a mixture of synchronous machines and inverter-based resources such as wind, solar, and energy storage. This transformation has led to a decrease in mechanical inertia, which necessitate a need for the new resources to provide frequency responses through their inverter interfaces. In this paper we proposed a new strategy based on model predictive control to determine the optimal active-power set-point for inverters in the event of a disturbance in the system. Our framework explicitly takes the hard constraints in power and energy into account, and we show that it is robust to measurement noise, limited communications and delay by using an observer to estimate the model mismatches in real-time. We demonstrate the proposed controller significantly outperforms an optimally tuned virtual synchronous machine on a standard 39-bus system under a number of scenarios. In turn, this implies optimized inverter-based resources can provide better frequency responses compared to conventional synchronous machines.
% resulted in a decrease in available rotational inertia which serves as an immediate response to frequency deviation due to disturbances. This has necessitated a need for more sophisticated controller structures to deal efficiently and effectively with this issue.
% In this work we propose a new control strategy called the Inverter Power Control (MIPC) which is capable of determining in real-time, the optimal active-power set-point for an IBR in the event of a disturbance. This control strategy utilizes a Model Predictive Control (MPC) optimization algorithm that relies on wide-area measurements as state inputs. This MIPC controller is capable of handling constraints in resources such as a limit on the power injection at every time step and/or a limit in the total energy available, by including this constraints into its optimization algorithm. It is also robust to noise and limited communication as it has a built in state and disturbance estimator. We compare the performance of this MIPC with that of an optimally design VSM by testing on the WSCC 3-Machine 9-bus test system and IEEE 39-bus test system to show the superior performance of the MIPC control.
\end{abstract}

% \begin{IEEEkeywords}
% IEEE, IEEEtran, journal, \LaTeX, paper, template.
% \end{IEEEkeywords}

% \IEEEpeerreviewmaketitle

\section{Introduction} \label{section:introduction}
The electric grid has been undergoing a transition from a network with dynamics fully governed by synchronous machines to a mixed-source network with dynamics governed by both synchronous machines and inverter-based resources (IBRs). This transition is marked by a reduction in the amount of mechanical inertia in the system, which has led to more pronounced frequency responses to disturbances and faults in the grid~\cite{machowski2011power,tielens2016relevance}. At the same time, by the virtue of the speed of power electronic circuits, IBRs such as solar, wind and energy storage have the capability to respond to frequency changes in the grid at a much faster rate than traditional generators with rotating masses. The challenge of how to best utilize these new capabilities has spurred much research interest in the last few years (e.g., see~\cite{milano2018foundations} and the references within). 

% in the grid at a much faster rate than traditional  but at the same time, an increase of resources  introduced new resources  The effect of this transition is a decline in the available inertia due to the rotating mass of the synchronous machines which can respond instantaneously to a deviation in frequency caused by a disturbance to the network by injecting active power \cite{machowski2011power}. This results in a larger increase in the frequency deviation, rate of change of frequency (ROCOF), and a slower settling time; all of which can lead to frequency instability in the network and an eventual collapse of the grid \cite{tielens2016relevance}. 

Various control strategies that utilizes the IBRs in providing frequency regulation services has been proposed. The goal of these strategies is to design the active power response of the IBRs to changes in frequency, such that some frequency response objective is minimized. For example, standard objectives of interests are the magnitude of the frequency deviation, the rate of change of frequency (ROCOF) and the settling time. A unique challenge in the control of IBRs is that they tend to face much tighter limits than conventional machines. For example, solar and wind resources cannot increase their power output beyond the maximum power tracking point, which introduces a hard (and asymmetrical) constraint on the action of the inverters. For a storage unit, it has only a limited amount of energy that can be used to respond to a disturbance.  
% as these IBRs have power electronic interfaces-\emph{inverters}-with potentials to react to disturbances and contribute to stabilizing and controlling the system frequency. To tap into these potentials, the inverters will require a sophisticated control and optimization structure with the ability to participate in controlling the frequency in a mixed-source network, effectively detect a disturbance by way of a deviation in frequency and optimally determine the amount of active power injection to stabilize the system.

Of the varying control strategies proposed for IBRs, \emph{Droop Control} \cite{van2015droop, simpson2013synchronization, ofir2018droop} and \emph{Virtual Synchronous Machines} (VSMs) \cite{van2009grid, sakimoto2011stabilization, zhong2011synchronverters} are the most popular as they function by mimicking the frequency-power dynamic response of a synchronous machine. As suggested by their names, droop control injects/absorbs an amount of active power in proportion to the frequency deviation, and VSM, in its basic configuration, acts as a second order oscillator to provide inertia and damping to the grid. The parameters (droop slope, inertia and damping constants) used in these strategies can be optimized using a number of techniques~\cite{AdemolaIdowu2018OptimalDO,borsche2017placement, poolla2017optimal}. 

%Due to the digital control nature of the VSM implementation, it can be further enhanced by incorporating virtual governors, virtual exciters and other power systems controllers~\cite{hirase2013grid, yao2017virtual}.

% The droop control injects/absorbs in addition to the reference power of the IBR, an active power proportional to the frequency deviation as determined by the droop gain coefficient while the VSMs power injection is proportional to the frequency deviation and ROCOF as determined by the virtual inertia and damping gain coefficients.

% To improve the performance of VSM
% There has been several efforts to improve the performance of the VSM: In \cite{AdemolaIdowu2018OptimalDO}, the virtual inertia and damping gain coefficients was optimally designed using the $\mathcal{H}_2$ norm as the objective function while in \cite{borsche2017placement, poolla2017optimal}, the optimal placement of IBRs through the virtual inertia and damping gain coefficients was considered. 

The structural simplicity of VSMs also leads to a fundamental limitation~\cite{AdemolaIdowu2018OptimalDO, wang2018adaptive}. Since there are only two parameters to tune (inertia and damping) in the basic VSM configuration, there is an inherent trade-off between different objectives and there is no choice of parameters that will make the frequency deviation, ROCOF, and settling time small at the same time~\cite{AdemolaIdowu2018OptimalDO}. While the performance of the VSM can be enhanced by incorporating virtual governors, virtual exciters and other power systems controllers in their virtual form~\cite{hirase2013grid, yao2017virtual}, it difficult to tune the multiple parameters of the combined virtual controller simultaneously, since the performance of one might negatively affect the other. In addition, it is also difficult to include hard constraints, since simply thresholding the output once the constraints are reached tend to lead to very poor performances~\cite{polycarpou2003line}. Adaptive rules can be used to alleviate this drawback somewhat, and works in \cite{wang2018adaptive, li2017self, zhang2018improved} change the parameter based on the measured frequency deviation and ROCOF values. However, it is difficult to find an optimal rule to update these parameters in real-time.

In this work, we propose a novel control strategy, based on model predictive control (MPC),  called the MPC-based Inverter Power Control (MIPC). We explicitly formulate the problem of finding the optimal active power set-point of an IBR to minimize the frequency deviation and the ROCOF. It turns out that this formulation also implicitly minimizes the systems settling time. More specifically, at any timestep, we simulate the dynamics of the systems for a finite horizon, then find the best set-points that optimizes the objective over that horizon. The first action is then adopted for the current timestep, and the process repeats.  Our approach is similar in spirit to the ones in~\cite{wang2018adaptive, li2017self, zhang2018improved} since an objective is optimized in an online fashion. However, instead of optimizing the parameters, we directly find the best power set-points. This approach turns out to provide both an easier optimization problem and better control performances. Namely, the hard constraints on the IBRs are explicitly included in the optimization process. 

A requirement of MPC is that the IBR must have a model of the system to be optimized. If wide-area measurements are available, then the system states can be obtained from these measurements~\cite{dorfler2014sparsity}. In some systems, only a limited buses are equipped with these measurement devices (e.g. PMUs). We show that our proposed MIPC framework is still applicable to these systems by building an observer to estimate unmeasured disturbances and states. Through simulation studies, we show that the MIPC strictly outperforms optimally tuned VSMs for the IEEE 39-bus system, even under limited communication and large measurement noises. 

This proposed controller finds practical application by enhancing the capability of the IBRs to participate in providing frequency regulation services. The additional power required can be obtained by running solar below its maximum power point to create sufficient headroom, utilizing the inertia from the decoupled rotating wind turbine and, leveraging on the stored energy in a battery. By explicitly considering hard constraints and costs on energy and power in the MPC formulation, economic considerations can be accounted for.

% which is capable of determining in real-time, the optimal real power set-point for an IBR in the event of a disturbance that minimizes the frequency deviation and ROCOF at each time step. This control strategy utilizes a Model Predictive Control (MPC) optimization algorithm that relies on wide-area measurements as state inputs to determine this real power set-point. This MIPC controller is capable of handling constraints in resources such as a limit on the power injection at every time step and/or a limit in the total energy available, by including the constraints into its optimization algorithm. It is also robust to noise and limited communication as it has a built-in state and disturbance estimator. We compare the performance of this MIPC with that of an optimally design VSM by testing on a 9-bus and a 39-bus system to show the superior and versatile performance of this proposed control strategy. 

The remainder of this paper is organized as follows: Section \ref{section:modeling} defines the models used in this paper. Section \ref{section:MIPC} presents the design and formulation of the MIPC algorithm. Section \ref{section:noise} presents the state and disturbance observer design. Section \ref{section:results} compares the performances of MIPC to VSMs in a standard test system. Section~\ref{section:conclusion} concludes the paper.

\section{Modeling} \label{section:modeling}
% \textit{Notations}:
We denote the real line by $\mathbb{R}$, the cardinality of a set $\mathcal{S}$ as $\vert \mathcal{S} \vert$, the $n \times n$ identity and zero matrices as $\vect{I}_n$ and $\vect{0}_n$, respectively. Matrices and vectors are denoted by a bold-faced variables.

\subsection{System Structure}
% The components of an electric power system can be broadly classified into two: power generating components and power consuming components. The power generating component in recent times is a combination of conventional synchronous generators with well studied dynamics and non-synchronous sources such as wind, solar and energy storage typically called inverter-based resources (IBRs) because they are connected via power electronics to the grid. The power consuming components consists of loads and lossy elements in the grid such as transmission lines. Loads are typically modeled using a constant model which is some variations in the algebraic power balance equation. The constant impedance load model is mostly used in dynamic simulations because it allows for combining the differential and algebraic system equations into differential equations only.

Steady state conditions in a power systems are achieved when there is a balance between the power produced by the generating sources and the power consumed by loads and lossy components. For stability analysis, the entire system can be reduced to an equivalent network via Kron reduction \cite{nishikawa2015comparative}. This eliminates passive and non-dynamic load buses and leaves only buses with at least one generating source connected. With this in place, frequency stability analysis can be carried out, with the frequency dynamics governed by the reactions of buses to active power imbalances in the system.

In this work, we assume the availability of state variables and network information for control purposes. In a later section, we will relax this assumption to partial availability of state variables from some generators.
% In modern power systems, wide area measurement system (WAMS) data, consisting of sensors and communication infrastructures, is increasingly available \cite{phadke2018phasor} as most balancing authorities require that new generators, greater than a certain MW, have phasor measurement units (PMU) installed \cite{pjm}. \bz{this contradicts the claim that limited communication is considered, move to MIPC section}

Because the generators and IBRs had different dynamics, we denote their sets by $\mathcal{G}$ and $\mathcal{I}$, respectively. Note that the total number of generating sources in the network is $\mathcal{N} := \mathcal{G} \cup \mathcal{I}$.

% We assume the existence of Wide Area Measurements (WAMS), consisting of sensors and communication infrastructures that can measure and send system state information to the inverter controller system. We focus on active power control in controlling the frequency response of the system.

\subsection{Synchronous Machines}
The rotor dynamics of each synchronous generator in a given power system is governed by the well-known swing equation \cite{sauer2017power}. Here we adopt a discretized version of the equations, which in per unit (p.u.) system is:
% This can be discretized and represented in per-unit (pu) form as:
\begin{equation} \label{eqn:swing_rotor_dis}
\begin{aligned}
\omega^{t+1}_i & = \omega^{t}_i + \frac{h}{m_i} \left(P_{\text{m},i}^t - P_{\text{e},i}^t - d_i \omega^{t}_i \right), \\
\delta^{t+1}_i & = \omega_{\text{b}} \left(\delta^{t}_i + h \ \omega^{t+1}_i\right),
\end{aligned}
\end{equation}
$\forall i \in \mathcal{G}$ where $h$ is the step size for the discrete simulation, $\delta_i$ (rad) is the rotor angle, $\omega = \bar{\omega}_i - \omega_0$ is the rotor speed deviation, $\omega_{\text{b}}$ is the base speed of the system, $m_i$ is the inertia constant, $d_i$ is the damping constant, $P_{\text{m},i}$ is the mechanical input power and $P_{\text{e},i}$ is the electric power output of the $i^{th}$ machine.

% With the assumption of a constant load impedance,
The electrical output power $P_{\text{e},i}$ is given by the AC power flow equation in terms of the internal emf $\vert E_i \vert$ and rotor angle $\delta_i$:
\begin{align} \label{eqn:pf_dis}
P_{\text{e},i}^t = \sum_{i \sim j} \vert E_i E_j \vert [g_{ij} \cos(\delta_i^t - \delta_j^t) + b_{ij}\sin(\delta_i^t - \delta_j^t)],
\end{align}
$\forall i,j \in \mathcal{G}$, where $g_{ij} + j b_{ij}$ is the reduced admittance between nodes $i$ and $j$. We assume the internal emf are constant because of the actions of the exciter systems.

The nonlinearity of the AC power flow in \eqref{eqn:pf_dis} makes \eqref{eqn:swing_rotor_dis} difficult to use for control applications.
Linearizing \eqref{eqn:swing_rotor_dis} around the nominal point and using the DC power flow approximation ~\cite{kundur1994power}, the bus dynamics become:
\begin{equation} \label{eqn:swing_lin_dis}
\begin{aligned}
\bigtriangleup  \omega^{t+1}_i & = \bigtriangleup  \omega^{t}_i + \frac{h}{m_i} \left(\bigtriangleup  P_{\text{m},i}^t - \bigtriangleup P_{\text{e},i}^t - d_i \bigtriangleup  \omega^{t}_i \right), \\
\bigtriangleup \delta^{t+1}_i & = \omega_{\text{b}} \left(\bigtriangleup \delta^{t}_i + h \ \bigtriangleup \omega^{t+1}_i \right),
\end{aligned}
\end{equation}
where $\bigtriangleup P_{\text{e},i}^t = \sum_{i \sim j} b_{ij} \bigtriangleup \delta^t_{ij}$ is the dc power flow between 2 buses. We model changes to the mechanical input power $\bigtriangleup P_{\text{m},i}^t$ by a combination of droop and automatic governor control (AGC) actions \cite{kundur1994power}.

% according to the discretized equation:
% \begin{align} \label{eqn:agc}
%     \bigtriangleup P_{\text{m},i}^{t+1} & = \frac{1}{1 + hk_1}\Big(\left[2 + hk - \frac{ h^2 k_2}{m_i}\right]\bigtriangleup P_{\text{m},i}^t - \bigtriangleup P_{\text{m},i}^{t-1} + \\ \nonumber
%     & h^2 \left[\left(\frac{k_2 d_i }{m_i} - k3\right) \bigtriangleup \omega_i^t  - \frac{k_2}{m_i} \left(\bigtriangleup P_{\text{e},i}^t\right) \right] \Big),
% \end{align}
% where $k_1$, $k_2$ and $k_3$ are the gain coefficients of the droop and AGC controller.

% $P_{I,i}$ in \eqref{eqn:swing_rotor_dis} models the impact of the inverter on the dynamics of the synchronous generators it is connected to and is typically designed to mimic the response of a synchronous machine to frequency deviation as in a Virtual Synchronous Machine (VSM) \cite{sakimoto2011stabilization}. This modifies \eqref{eqn:swing_rotor_dis} by replacing $P_{I,i}$ as an additional machine. This is discussed further in the next chapter.

\subsection{Virtual Synchronous Machine (VSM)} \label{subsection:vsm}
% \subsubsection{Model and Control} \label{subsubsection:inverter control}
From the network point of view, the grid-connected IBR is seen as producing a constant power according to its predetermined set-point and fast dynamics governed by closed-loop controls actions \cite{pico2019transient}. When configured in the grid-following mode, these controls help maintain the output power of the IBRs while remaining synchronized to the terminal voltage set by the grid\cite{rocabert2012control}. For system analysis, the inverter can be modeled as a voltage source behind a reactance, much like a synchronous machine.
% as shown in Fig. \ref{fig:vsc}.
% \begin{figure}[hbpt]
% \centering
% \includegraphics[width = 0.8\columnwidth]{VSC.pdf}
% \caption{Schematic diagram of a grid-connected IBR showing its configuration and representation as a voltage source behind a reactance }
% \label{fig:vsc}
% \end{figure}

In the event of a power imbalance in the network reflected by a frequency deviation, an inverter does not have a "natural" response to frequency deviation as synchronous machines does since they are made of power electronics components and have no rotating mass. To elicit some response, an additional control loop is therefore needed to enable the inverters to participate in frequency control by changing the power set-point of the inverter based on frequency measurements. The concept of virtual synchronous machine (VSM) has been proposed in literature to provide this additional control loop and it comes in different configurations~\cite{sakimoto2011stabilization,alipoor2015power,bevrani2014virtual}. The basic idea is to mimic the behavior of a synchronous machine's response to a frequency deviation by choosing appropriate gains corresponding to the inertia and damping of the machines and producing power proportional to the ROCOF and frequency deviation. Since the response of the inverter is entirely digital, it can be programmed with almost arbitrary functions. 
% In this work, we adopt the VSM configuration in \cite{bevrani2014virtual} where the additional control loop is as shown in Fig. \ref{fig:vsm}.
% \begin{figure}[hbpt]
%     \centering
%     \includegraphics[width = 0.8\columnwidth]{./Images/update/VSM_block.pdf}
%     \caption{Block Diagram showing the operation of the VSM control loop which utilizes a feedback PD control mechanism to enable the IBR participate in frequency control}
%     \label{fig:vsm}
% \end{figure}
% Under this VSM configuration, the IBR control measures the grid frequency denoted as $\bigtriangleup \omega_{ibr}$ using the phase locked loop (PLL) \cite{yazdani2010voltage} and computes the additional power 

In this work, we adopt the VSM configuration in \cite{bevrani2014virtual}, where the additional power required to combat a frequency deviation is computed using:
\begin{equation} \label{eqn:p_add}
    \begin{aligned}
\bigtriangleup P = \bigtriangleup P_{\text{km}} + \bigtriangleup P_{\text{kd}}= K_\text{m} \frac{d \bigtriangleup \omega_{\text{ibr}}}{dt} + K_\text{d} \bigtriangleup \omega_{\text{ibr}}.
\end{aligned}
\end{equation}
The local frequency at the IBR node $\bigtriangleup \omega_{\text{ibr}}$ is approximated by the center of inertia (COI) frequency \cite{pierre2019bulk, pico2019transient}, which is an inertia-weighted average frequency given by:
\begin{equation} \label{eqn:local_freq}
    \begin{aligned}
        \bigtriangleup \omega_{ibr} = \frac{\sum_{i=1}^n m_i \bigtriangleup \omega_i}{\sum_{i=1}^n m_i}
    \end{aligned}
\end{equation}
where $n = \vert \mathcal{G} \vert$, $\bigtriangleup \omega_i$ is the rotor speed deviation, and $m_i$ is the inertia constant of the $i^{th}$ synchronous generators in the network. The gains $K_\text{m}$ and $K_\text{d}$ in \eqref{eqn:p_add} represent the virtual inertia and damping constants respectively. In contrast to synchronous machines where the constants are decided by the physical parameters, these constants of the VSM can be optimized over~\cite{AdemolaIdowu2018OptimalDO}.

In the next section, we fully leverage the flexibility of the power electronic interfaces using a MPC framework.

\section{MPC-based Inverter Power Control (MIPC)} \label{section:MIPC}
\begin{figure}[hptb]
\centering
\includegraphics[width = \columnwidth]{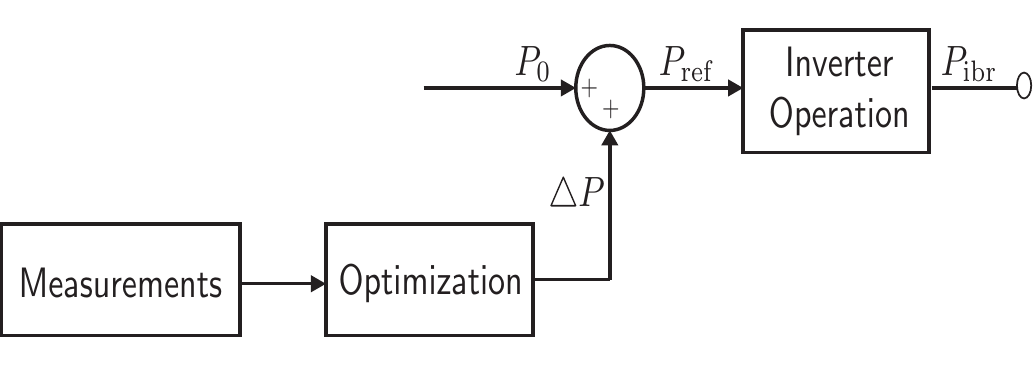}
\caption{Block Diagram showing the operation of the MIPC controller which utilizes state and network variables to modify the real power set-point of the IBR at each timestep} 
\label{fig:MIPC}
\end{figure}

In this work, we propose a novel method for controlling the output power of the IBR, called the Inverter Power Control (MIPC). This controller functions by modifying the initial real power set-point $P_{\text{0}}$ to a new set-point $P_{\text{ref}}$ as shown in Fig. \ref{fig:MIPC} at each time step such that a weighted sum of the frequency deviation and ROCOF is minimized. Due to the timescale difference between IBRs and synchronous machines, the real power set-points of an IBR can be set almost instantaneously. Therefore, the important question becomes how to solve the optimization problem at each time step fast enough to find the real power set-point and how much communication is required in performing these calculations. In this section, we describe how to formulate the optimization problem and provide an efficient algorithm, assuming all of the information are known at the IBR. The next section then discusses how to deal with limited and noisy measurements, as well as incomplete communication. 

For the $k^\text{th}$ IBR, let $u_k$ denote its angle (referenced to the slack-bus). We think of this $u_k$ as the control variable in the optimization problem. Note that the actual control of the IBR is not done via angle control, rather, we use the optimized $u_k$ to find the corresponding active power output of the inverter, then set the inverter to that power. 
% the We define a variable $u_k \coloneqq \delta_{k \in \mathcal{I}}$ and note that $u_k$ is not the actual inverter angle as that cannot be changed since it is set by the current control loop of the inverter. We instead use $u_k$ as a means to determine the optimal power output from the inverter in a soon to be defined optimization problem.
To determine this real power set-point at a given time step, consider the power flow equation in \eqref{eqn:pf_dis}, we write out the $i^{\text{th}}$ generator's power output $P_{\text{e},i}$ into two parts: power flowing from the $i^{\text{th}}$ generator to other generator denoted as $P_{\text{eG},i}$ and from the $i^{\text{th}}$ generator to IBRs denoted as $P_{\text{eI},i}$, such that:
\begin{equation} \label{eqn:gen_inv_power}
    \begin{aligned}
      P_{\text{e},i} & = P_{\text{eG},i} + P_{\text{eI},i} \\
        & = \sum_{i \sim j, i,j \in \mathcal{G}} \vert E_i E_j\vert [g_{ij} \cos(\delta_i - \delta_j) + b_{ij}\sin(\delta_i - \delta_j)] \\ 
& + \sum_{i \sim k, i,\in \mathcal{G}, k \in \mathcal{I}} \vert E_i E_k \vert [g_{ik} \cos(\delta_i - u_k) + b_{ij}\sin(\delta_i - u_k)],
    \end{aligned}
\end{equation}
and the output power from the $k^{\text{th}}$ IBR denoted as $P_{\text{ibr},k}$ can also be written in two parts as: 
\begin{equation} \label{eqn:tot_inv_power}
\begin{aligned}
P_{\text{ibr},k} & = P_{\text{ibr},ki} + P_{\text{ibr},kj}, \\
& = \sum_{k \sim i, k \in \mathcal{I}, i \in \mathcal{G}} \vert E_k E_i\vert [g_{ki} \cos(u_k - \delta_i) + b_{ki}\sin(u_k - \delta_i)] \\ 
& + \sum_{k \sim j, j, k \in \mathcal{I}} \vert E_k E_j \vert [g_{kj} \cos(u_k - u_j) + b_{kj}\sin(u_k - u_j)].
\end{aligned}
\end{equation} 
% In this form, the IBR does not need to be written in the synchronous machine form as with the VSM structure and requires only one control variable per IBR instead of the optimal selection of virtual inertia and damping gains.

\subsection{Nonlinear Optimization Problem}
% To respond to a disturbance, the IBR's active power output (or the angle $u^t$ in \eqref{eqn:inv_power}) can be optimized at each time step. 
% \bz{Disturbances should be present in this equation already, probably in 8(c). Remember, this is the full formulation, so other two subsections cannot add any new variables.}
At any timestep, we consider the behavior of the system $N$ steps ahead. Without loss of generality, we start the problem at time $t=0$. The control variables are the inverter angles, which we denote as $\vect{u}^0,\vect{u}^1,\dots,\vect{u}^{N-1}$. Once these are set, the rest of the system are governed by their swing equations. As stated before, the objective is to minimize a function of the frequency deviation and the ROCOF, and the MIPC problem is given by:
% As described in section \ref{section:introduction}, the important frequency response metrics are the frequency deviation and ROCOF which are used in measuring a network's stability, and the natural objective is then to minimize these quantities. 
% For an effective and efficient response to a transient disturbance, the inverter angle $u^t$ in \eqref{eqn:inv_power} has to be optimally determined at every time step, leading to an optimal injection of power by the inverter. As described in section \ref{section:introduction}, the important frequency response metrics are the frequency deviation and ROCOF which are used in determining the network's frequency stability. The objective is to keep these values as small as possible especially the frequency deviation which has to be within the $59.5-60.5$Hz range in most interconnections.
% This objective coupled with the network dynamics and resource constraints can be written in the form of an optimization  problem to be carried out by the MIPC as:
\begin{subequations} \label{eqn:opt_alg}
\begin{align} 
& \underset{\{\vect{u}^0, \vect{u}^1, \dots, \vect{u}^{N-1}\}}{\text{Min.}} \ \sum_{t=0}^{N-1}
\left\{\Vert \vect{\omega}^{t+1} \Vert^2_2 \ + \frac{1}{h}\Vert \vect{\omega}^{t+1} - \vect{\omega}^t \Vert^2_2\right\}\\
& \text{s.t.}
\label{eqn:speed}
\ \omega^{t+1}_i= \omega^t_i + \frac{h}{m_i} \big(P_{\text{m},i}^t - P_{\text{e},i}^t  - d_i \omega^t_i  - \bigtriangleup P_i^t \big), \ \forall i \in \mathcal{G}\\
\label{eqn:Pe}
& P_{e,i}^t = \text{Equation} \ \eqref{eqn:gen_inv_power}, \ \forall i \in \mathcal{G}\\
% & P_{I,i}^t = \sum_{k\in \mathcal{I}} G_{ik} \cos(\delta_i^t - u_k^t) + B_{ij}\sin(\delta_i^t - u_k^t), \forall i \in \mathcal{G}\\
% \label{eqn:PI}
% & P_{I,i}^t =  \text{Equation} \ \eqref{eqn:inv_power}, \ \forall i \in \mathcal{G}\\
\label{eqn:Pinv}
& P_{ibr,k}^t = \text{Equation} \ \eqref{eqn:tot_inv_power}, \ \forall k \in \mathcal{I}\\
% & P_{I,i} = \sum_{i \sim j, i \in \mathcal{G}, j \in \mathcal{I}} \vert E_i E_j\vert [g_{ij} \cos(\delta_i - u_j) + b_{ij}\sin(\delta_i - u_j)] \\ 
% & P_{I,i} = \sum_{i \sim j, i,j \in \mathcal{I}} \vert E_i E_j \vert [g_{ij} \cos(u_i - u_j) + b_{ij}\sin(u_i - u_j)] \\
\label{eqn:max_power}
&P_{\text{ibr,min},k}^t \leq P_{\text{ibr},k}^t \leq P_{\text{ibr,max},k}^t, \\
\label{eqn:max_energy}
% & E_{\text{ibr, min},k} \leq \sum^T P^t_{\text{ibr,k}} t \leq E_{\text{ibr, max},k},
& \sum_{t=1}^N P^t_{\text{ibr,k}} \leq E^t_{\text{ibr, tot},k},
\end{align}
\end{subequations}
where $\vect{\omega}^{t+1} \in \mathbb{R}^{\mathcal{|G|}}$ is a vector of all machine frequency deviations at the next time step and $\vect{\omega}^{t+1} - \vect{\omega}^t$ is a vector of all machine ROCOF between the current and next time step. The evolution of $\vect{\omega}$ is given in \eqref{eqn:speed} (swing equations) with the added $\bigtriangleup P_i$ used to denote disturbances to the network which can be either a loss in generation or load, the power constraints are given in \eqref{eqn:max_power} and the energy constraints are in \eqref{eqn:max_energy}. Here we take the frequency deviation and the ROCOF to be equally weighted for simplicity, but their weighting can be adjusted as needed for different practical scenarios. 

After \eqref{eqn:opt_alg} is solved, the control variable $\vect{u}^{0}$ is substituted into the power flow equations \eqref{eqn:tot_inv_power} to find the active power set-points of the IBRs. Then the IBRs hold their power at these set-points until the next time the optimization problem is solved. This is in line with the MPC convention, where the optimal control action is computed for the whole control horizon and only the first action is used. The process is then repeated again to determine the new control action.

It turns out that the AC power flow equations in \eqref{eqn:gen_inv_power} and \eqref{eqn:tot_inv_power} makes the problem nonlinear and difficult to solve in real-time. Therefore, the next two sections uses DC power flow to obtain an approximate problem that is much easier to solve. 

% At every time step, the optimal control variable $\vect{u}^t$ can be determined by solving the optimization algorithm in \eqref{eqn:opt_alg} for a finite-time horizon $N$ using \emph{Nonlinear Model Predictive Control} (NMPC). As with the MPC algorithm, the first step control is implemented and the process repeated again.

\subsection{Unconstrained Linearized Problem} \label{subsection:unconstrained}
% The non-linearity of optimization algorithm in \eqref{eqn:opt_alg} makes an on-line implementation difficult. Since we are repeatedly solving \eqref{eqn:opt_alg} in every time step and only taking the first action, it would be beneficial to replace \eqref{eqn:opt_alg} with a tractable approximation that can be solved efficiently. 
The main source of non-linearity in \eqref{eqn:opt_alg} comes from the AC power flow equations in \eqref{eqn:Pe} and \eqref{eqn:Pinv} and we use the standard DC power flow model from \eqref{eqn:swing_lin_dis} to approximate these equations. 

Therefore, at bus $i \in \mathcal{G}$  (synchronous machines), we have:
% \begin{align}
% \label{eqn:pe}
% \bigtriangleup P_{eG, i} & = \sum_{i \sim j, j \in \mathcal{G}} b_{ij}(\bigtriangleup \delta_i - \bigtriangleup \delta_j), \\
% \label{eqn:pi}
% \bigtriangleup P_{eI, i} & = \sum_{i \sim k, k \in \mathcal{I}} b_{ik}(\bigtriangleup \delta_i - u_k),
% \end{align}
% therefore, 
\begin{equation}
\begin{aligned}
\bigtriangleup & P_{e,i} = \bigtriangleup P_{eG, i} + \bigtriangleup P_{eI, i}   \\ 
 &= \sum_{i \sim j, j \in \mathcal{G}} b_{ij} (\bigtriangleup \delta_i - \bigtriangleup \delta_j) + \sum_{i \sim j, k \in \mathcal{I}}   b_{ik}(\bigtriangleup \delta_i - u_k), \label{eqn:p_tot_mach}
% & = (\sum_{i \sim j, j \in  \mathcal{N}}b_{ij}) \bigtriangleup \delta_i  + \sum_{i \sim j, j \in  \mathcal{G}} -b_{ij} \bigtriangleup \delta_j + \sum_{i \sim j, k \in \mathcal{I}} -b_{ik}u_k.
\end{aligned}    
\end{equation}
which can be written in matrix form as:
\begin{equation}
\begin{aligned}
\bigtriangleup P_{e} 
& = 
\underbrace{
\begin{bmatrix}
b_{ii} & -b_{ij}\\
-b_{ji}& b_{jj} 
\end{bmatrix}
}_{\vect{B}_{\text{GG}}}
\begin{bmatrix}
\bigtriangleup \delta_i\\
\bigtriangleup \delta_j
\end{bmatrix}
+
\underbrace{
\begin{bmatrix}
-b_{ik}\\
-b_{jk}
\end{bmatrix}
}_{\vect{B}_{\text{GI}}}
u_k,
% & \forall \ i \neq j, \ i \neq k, \ i,j \in \mathcal{G} \ \text{and} \ k \in \mathcal{I}
\end{aligned}
\end{equation}
where $\vect{B}_{\text{GG}}$ contains the connection between synchronous generators and $\vect{B}_{\text{GI}}$ contains the connection between a synchronous generator and IBRs. In state space form, it becomes
\begin{align} \label{eqn:swing_ss_dis}
\vect{
\underbrace{
\begin{bmatrix} 
\bigtriangleup \omega^{t+1}\\
\bigtriangleup  \delta^{t+1}
\end{bmatrix}
}_{x^{t+1}}}
= 
\vect{
\underbrace{
\begin{bmatrix}
- M^{-1}D & - M^{-1}B_{\text{GG}}\\
I_n & 0_n
\end{bmatrix}
}_{\bar{A}}
\underbrace{
\begin{bmatrix} 
\bigtriangleup \omega^t\\
\bigtriangleup \delta^t
\end{bmatrix}
}_{x^t}}\\ \nonumber
+
\vect{
\underbrace{
\begin{bmatrix} 
-M^{-1}B_{\text{GI}}\\
0_n
\end{bmatrix}
}_{\bar{B}_u}
u^t
+
\underbrace{
\begin{bmatrix} 
-M^{-1}\\
0_n
\end{bmatrix}
}_{\bar{B}_d}
\underbrace{
\bigtriangleup P^t
}_{d^t}
}
\end{align}
where $\vect{\bigtriangleup  \delta} \in \mathbb{R}^n$ is the rotor angles deviation, $\vect{\bigtriangleup \omega} \in \mathbb{R}^n$ is the rotor speed deviation, $\vect{M} = \text{diag}(m_1,\dots,m_n) \in \mathbb{R}^{n \times n}$, $\vect{D} = \text{diag}(d_1,\dots,d_n) \in \mathbb{R}^{n \times n}$, $\vect{\bigtriangleup P^t} \in \mathbb{R}^n$ is vector of all power deviations which comes from the disturbances and noises in the system, denoted by $\vect{d}^t$. 
% disturbances and governor responses denoted as $\vect{d^t}$.
% where $\vect{\bigtriangleup  \delta} \in \mathcal{R}^n$ is the rotor angles deviation, $\vect{\bigtriangleup \omega} \in \mathcal{R}^n$ is the rotor speed deviation, $\vect{M} = diag(m_1,\dots,m_n) \in \mathcal{R}^{n \times n}$ where $m_i$ is the moment of inertia at bus $i$, $\vect{D} = diag(d_1,\dots,d_n) \in \mathcal{R}^{n \times n} \in \mathcal{R}^n$ where $d_i$ is the damping at bus $i$, $\vect{P_m} \in \mathcal{R}^n$ is vector of all disturbances at the buses the mechanical input power, and $\vect{P_{inv}} \in \mathcal{R}^n$ is the vector of all inverter output power.\\

Since the MIPC does not know the disturbance or noise impacting the system, we use a two step process to solve the optimization problem. First, we ignore the disturbance term such that the MIPC's model of the system is:

% it makes use of only the known portion of the model, that is:
\begin{align} \label{eqn:swing_ss_red}
\vect{
\underbrace{
\begin{bmatrix} 
\bigtriangleup \omega^{t+1}\\
\bigtriangleup  \delta^{t+1}
\end{bmatrix}
}_{x^{t+1}}}
= 
\vect{
\underbrace{
\begin{bmatrix}
- M^{-1}D & - M^{-1}B_{GG}\\
I_n & 0_n
\end{bmatrix}
}_{\bar{A}}
\underbrace{
\begin{bmatrix} 
\bigtriangleup \omega^t\\
\bigtriangleup \delta^t
\end{bmatrix}
}_{x^t}}\\ \nonumber
+
\vect{
\underbrace{
\begin{bmatrix} 
-M^{-1}B_{GI}\\
0_n
\end{bmatrix}
}_{\bar{B}}
u^t}.
\end{align}
Note that in this case the MIPC's model of the system is actually wrong since the disturbances are not modeled. It turns out that this model is still useful, since the measurements are updated every time the MPC problem is solved, and this compensates for using a wrong model. In the rest of this section, we focus on solving the optimization problem using the model in \eqref{eqn:swing_ss_red} since it illustrates our methodology. Of course, when the measurement noise in the system is large or not every bus is equipped with wide-area measurement devices, it becomes necessary to explicitly estimate the mismatch between the model and the actual system. We do so in section \ref{section:noise}. 
% we will revisit the model in \eqref{eqn:swing_ss_dis} and propose a way of estimating $\vect{d^t}$ from the received state measurements.

We reformulate the objective function in terms of the network model in \eqref{eqn:swing_ss_red} by defining the output of the linearized model as the frequency deviation $\vect{\bigtriangleup \omega^t}$:
\begin{align}
\vect{y^t = 
\underbrace{
\begin{bmatrix}
I_n & 0_n
\end{bmatrix}
}_{C}
x^t    
=
\bigtriangleup \omega^t}
\end{align}
such that the ROCOF becomes:
\begin{align}
    \vect{\bigtriangleup y^t} = \frac{1}{h} \left[ \vect{y^t - y^{t-1}}\right] 
    = \frac{1}{h} \left[ \vect{\bigtriangleup \omega^t - \bigtriangleup \omega^{t-1}}\right]
\end{align}

% Optimization problem can be solved using typical LQR/MPC. Only choose to minimize state deviation since we assume limitless storage/ state deviation is priority and typically the objective is competing.
The MIPC optimization algorithm in \eqref{eqn:opt_alg} without the power limit constraint \eqref{eqn:max_power} and total energy constraints \eqref{eqn:max_energy} can now be written as a linear quadratic programming problem:
\begin{equation} \label{eqn:opt_inv}
\begin{aligned}
& \underset{\vect{u^t}}{\text{Min.}} \
J = \frac{1}{2} \sum_{t = 0}^{N-1} \left[ \vect{y^{t^T} Q_1 y^t} + \vect{\bigtriangleup y^{t^T} Q_2 \bigtriangleup y^t} \right]\\ 
& \text{s.t.}
\quad \vect{x^{t+1} = \bar{A} x^t +  \bar{B} u^t}\\
& \qquad \quad \  \vect{y^t  = C x^t}. 
\end{aligned}
\end{equation}

This can be written in matrix form for $N$ time step ahead as:
\begin{equation} \label{eqn:opt_lp}
\begin{aligned} 
& \underset{\vect{u}}{\text{Min.}} \
J = \frac{1}{2}\vect{{x^0}^T G x^0 + {x^0}^T F u } + \frac{1}{2}\vect{u^T H u}. 
\end{aligned}
\end{equation}
where $\vect{H}$ and $\vect{F}$ are constant matrices depending on $\vect{\bar{A}}$ and $\vect{\bar{B}}$ (see Appendix).
% Not surprisingly, the optimal solution to this unconstrained problem is linear in the starting point $\vect{x}^0$:
% % The solution to this unconstrained MIPC optimization in \eqref{eqn:opt_inv} for $N$ time steps ahead, can be obtained in closed form as:
% \begin{align}\label{eqn:opt_closed}
%     \vect{u^*} = \vect{-H^{-1} F^T x^0}.
% \end{align}
% This solution can be interpreted as a linear policy, where the optimal action is determined as a linear function of the current state information. 

\subsection{Constrained Linearized Optimization Problem}
In the presence of constraints, the problem becomes:
\begin{equation} \label{eqn:opt_qp}
\begin{aligned} 
& \underset{\vect{u}}{\text{Min.}} \
J = \frac{1}{2}\vect{{x^0}^T G x^0 + {x^0}^T F u } + \frac{1}{2}\vect{u^T H u} \\
& \text{s.t.}
\quad \vect{L u \leq W + V x^0},\\ 
\end{aligned}
\end{equation}
where $\textbf{L}, \vect{W}$ and $\vect{V}$ encodes the constraints. In this paper, we consider  constraints on the power output at each time step \eqref{eqn:max_power} and constraints on the total energy available \eqref{eqn:max_energy}. This problem is a linearly constrained quadratic program and is convex. 

\subsubsection{Power Output Constraint}
In practical considerations, there can be a limit on the amount of instantaneous power that can be drawn from the IBR due to factors such as the distance to the maximum power tracking operating point, the current ratings and switching speed of some power electronics components, and also power capability or C-rate of a battery. 

The transformation of the minimum and maximum instantaneous power limit from  $P_{\text{ibr,min},k}^t \leq P_{\text{ibr},k}^t \leq P_{\text{ibr,max},k}^t$ to the linear constraint in \eqref{eqn:opt_qp} involves writing the linearized power output of the $k^{\text{th}}$ IBR at time step $t$ in terms of the control variable $u^t$ and states $x^t$, and then stacking them in matrix form for the $N$ control horizon. 

The linearized output power $P_{\text{ibr},k}^t$ is written in terms of the power flow to generators and to other IBRs as:
\begin{equation*}
\begin{aligned}
 P_{\text{ibr},k}^t  &= \sum_{k \sim i, k \in \mathcal{I}, i \in \mathcal{G}} b_{ki}(u_k^t - \bigtriangleup  \delta_i^t) + \sum_{k \sim j, j, k \in \mathcal{I}} b_{kj}(u_k^t - u_j^t) \\ 
& = -\sum_{k \sim i, i \in \mathcal{G}}b_{ki} \bigtriangleup \delta_i^t + \sum_{k \in \mathcal{I}} b_{kk} u_k^t  - \sum_{k \sim j, j \in \mathcal{I}}b_{kj} u_j^t ,
\end{aligned}    
\end{equation*}
which can be written in matrix form as:
\begin{equation} \label{eqn:lin_ibr_power}
 P_{\text{ibr},k}^t =
\vect{
\begin{bmatrix}
0_n & -[B]_{ki} 
\end{bmatrix}
x^t
+
\begin{bmatrix}
     -[B]_{kj} &  [B]_{kk}
\end{bmatrix}
u^t}
\end{equation}

Stacking \eqref{eqn:lin_ibr_power} for a $N$ time horizon and writing the linear system dynamics in term of the initial state results in a form:
\begin{align} \label{eqn:inv_pow_limit}
\vect{
P_{\text{ibr},k} 
=
B_{\text{p1}} x^0 + B_{\text{p2}} u},
\end{align}
which can finally be written in the linear constraint form of \eqref{eqn:opt_qp} as
\begin{align}\label{eqn:opt_qp_power} 
\vect{
\underbrace{
\begin{bmatrix} 
-B_{\text{p2}} \\
B_{\text{p2}} 
\end{bmatrix}
}_{L}
\vect{u} 
\leq
\underbrace{
\begin{bmatrix} 
-P_{\text{ibr, min}}\\
P_{\text{ibr, max}}
\end{bmatrix}
}_{W}
+
\underbrace{
\begin{bmatrix} 
B_{\text{p1}}\\
- B_{\text{p1}} 
\end{bmatrix}
}_{V} x^0.
}
\end{align}
See appendix for derivation.

\subsubsection{Total Energy Constraint} 
This constraint occurs when there is a limit on the energy capacity of the IBR as in the case of a battery. For this constraint to be fully satisfied, the total energy not only at the end of the control
horizon but also at each rolling sum of the consecutive time step should be less than the maximum energy capacity.

As with the power output constraint, the total energy constraint $\sum_{t=1}^N P^t_{\text{ibr,k}} \leq E^t_{\text{ibr, tot},k}$ can also be written in the linear constraint form in \eqref{eqn:opt_qp} by taking the rolling sum over the inverter power output matrix in \eqref{eqn:inv_pow_limit}. This results in another matrix of the form:
\begin{align} \label{eqn:inv_ener_limit}
\vect{
E_{\text{ibr},k} 
=
B_{\text{e1}} x^0 + B_{\text{e2}} u}.
\end{align}
To avoid a sudden decline in the power output when the maximum available energy limit is reached, a rate constraint can be added to the power output decline between a specified consecutive time step. This can also be represent in the form of \eqref{eqn:opt_qp} by taking a one time step difference of the IBR power output matrix in \eqref{eqn:inv_pow_limit}, that is, a difference between the next time step and current time step IBR power output. This results in a matrix of the form:
\begin{align} \label{eqn:inv_rate_limit}
\vect{
B_{\text{r1}} x^0 + B_{\text{r2}} u}
\leq
\bigtriangleup P_{\text{ibr},k} 
\triangleq \epsilon.
\end{align}
Equation \eqref{eqn:inv_ener_limit} and \eqref{eqn:inv_rate_limit} can finally be written in the linear constraint form of \eqref{eqn:opt_qp} as
\begin{align}\label{eqn:opt_qp_energy} 
\vect{
\underbrace{
\begin{bmatrix} 
B_{\text{e2}}\\
B_{\text{r2}}
\end{bmatrix}
}_{L}
\vect{u} 
\leq
\underbrace{
\begin{bmatrix} 
E_{\text{ibr, tot}}\\
\epsilon
\end{bmatrix}
}_{W}
+
\underbrace{
\begin{bmatrix} 
- B_{\text{e1}} \\
- B_{\text{r1}}
\end{bmatrix}
}_{V} x^0.
}
\end{align}
where $\vect{\epsilon}$ is a vector of IBR power output rate limit for each one time step difference. See appendix for derivation.

Even with constraints, a linear quadratic program can be solved extremely efficiently for systems with thousands of variables and constraints~\cite{nocedal2006numerical}. Again, to actually implement the controller, we compute and set the power output of the IBRs. 

% \section{Effect of Disturbance/Noise} \label{section:noise}

% \section{MIPC with State and Disturbance Estimation} \label{section:noise}
\section{MIPC with State Estimation and Limited Communication} \label{section:noise}
In section \ref{section:MIPC}, the MIPC controller was designed using the reduced linearized model of the network as in \eqref{eqn:swing_ss_red} and under the assumption of a full state measurement. When operating this controller in a realistic setting, we would want the controller to be robust against issues such as model mismatch, that is, the difference between the actual system model and the linearized model used by the MIPC; noisy measurements, and incomplete measurements because of limited communication between buses. 

\begin{figure}[hbpt]
\centering
\includegraphics[scale = 1.1]{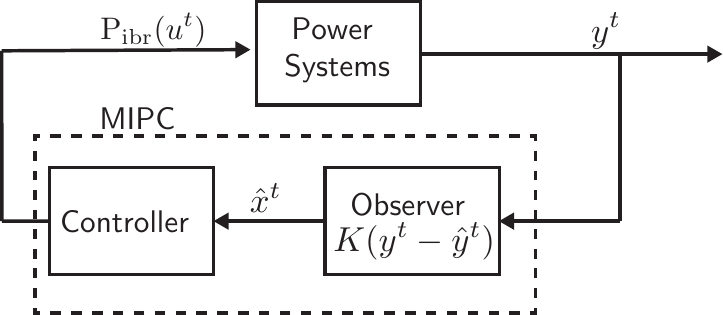}
\caption{Observer integrated MIPC in a power systems. The state measurements received by the observer in addition to the predicted state by the model is used to estimate the true state of the system.} 
\label{fig:observer}
\end{figure}

We address these issues in this section by integrating an observer into the MIPC controller system according to Fig. \ref{fig:observer} to enable the controller estimate a better model of the system from the received measurements.

Let the dynamics of the actual power systems governed by \eqref{eqn:swing_rotor_dis} be represented concisely as:
\begin{equation}\label{eqn:actual_system}
\begin{aligned}
\vect{x^{t+1}} & = \vect{f(x^t, u^t)}\\
\vect{y^{t}} & = \vect{g(x^t, u^t)}.
\end{aligned}
\end{equation}
A simple discrete observer model design \cite{pannocchia2003disturbance} for the system in \eqref{eqn:actual_system} can be written as: 
\begin{equation} \label{eqn:observer_model}
\begin{aligned}
\vect{\hat{x}^{t|t}} & = \vect{\hat{x}^{t|t-1} + K(y^t - \hat{y}^t)}\\
\vect{\hat{y}^{t}} & = \vect{C\hat{x}^{t|t-1}},
\end{aligned}
\end{equation}
% \begin{equation} \label{eqn:observer_model}
% \begin{aligned}
% \vect{\hat{x}^{t+1}} & = \vect{A\hat{x}^t + K(y^t - \hat{y}^t)+ Bu^t}\\
% \vect{\hat{y}^{t}} & = \vect{C\hat{x}^t},
% \end{aligned}
% \end{equation}
% where the variables with $\vect{\hat{x}}$ and $\vect{\hat{y}}$ are the observer estimated state and output variable, 
where the notation $\vect{\hat{x}}^{t|\tau}$ means the prediction of $\vect{x}^t$ made at time $\tau$. Therefore the variables with $\vect{\hat{x}^{t|t}}$ is the updated observer state prediction 
based on new measurement $\vect{y^t}$, $\vect{\hat{x}^{t|t-1}}$ is the observer state prediction of the current time step using measurements from the previous time step, and $K$ is a gain chosen such that the error between the measured and predicted state $\vect{y^t - \hat{y}}^t$ is quickly driven to zero.

\subsection{State and Disturbance Estimation}
To estimate the state and disturbance in a noisy system with model mismatch and other forms of disturbance, we redefine $\vect{d^t}$ as a vector of all disturbances. We then integrate an input/output constant disturbance model \cite{pannocchia2003disturbance} into the MIPC system model in \eqref{eqn:swing_ss_dis} to obtain:

\begin{equation} \label{eqn:observer_plant}
\begin{aligned}
\vect{\hat{x}^{t+1}} & = \vect{\bar{A}\hat{x}^t + \bar{B}_u u^t + \bar{B}_d \hat{d}^t}\\
\vect{\hat{d}^{t+1}} & = \vect{\hat{d}^t} \\
\vect{\hat{y}^t} & = \vect{C \hat{x}^t + C_d \hat{d}^t},
\end{aligned}
\end{equation}
where the disturbance $\vect{\hat{d}}$ is modeled as a constant disturbance for the control period. Equation \eqref{eqn:observer_plant} can then be written in an augmented form as: 
% where the estimated states, disturbance and output are denoted as $\vect{\hat{x}^t, \hat{d}^t}$ and $\vect{\hat{y}^t}$ respectively to distinguish them from the actual states, disturbance and output $\vect{x^t, d^t}$ and $\vect{y^t}$ respectively. 
\begin{align} \label{eqn:swing_ss_dis_aug}
\vect{
\begin{bmatrix} 
\hat{x}^{t+1}\\
\hat{d}^{t+1}
\end{bmatrix} }
% \underbrace{
% \begin{bmatrix} 
% \hat{x}^{t+1}\\
% \hat{d}^{t+1}
% \end{bmatrix}
% }_{z^{t+1}} }
& = 
\vect{
\begin{bmatrix}
\bar{A} & \bar{B}_d\\
0 & I
\end{bmatrix}
% \underbrace{
% \begin{bmatrix}
% \bar{A} & \bar{B}_d\\
% 0 & I
% \end{bmatrix}
% }_{\hat{A}}
\begin{bmatrix} 
\hat{x}^t\\
\hat{d}^t
\end{bmatrix}
% \underbrace{
% \begin{bmatrix} 
% \hat{x}^t\\
% \hat{d}^t
% \end{bmatrix}
% }_{z^t}
+
\begin{bmatrix} 
\bar{B}_u\\
0
\end{bmatrix} u^t} \\ \nonumber
% \underbrace{
% \begin{bmatrix} 
% \bar{B}_u\\
% 0
% \end{bmatrix}
% }_{\hat{B}}
% u^t} \\ \nonumber
\vect{\hat{y}^t}
& = 
\vect{
\begin{bmatrix}
C & C_d    
\end{bmatrix}
\begin{bmatrix} 
\hat{x}^t\\
\hat{d}^t
\end{bmatrix} }.
\end{align}
The predicted augmented state and disturbance can then be estimated using the observer model in \eqref{eqn:observer_model} as:
\begin{align} \label{eqn:observer_est}
\vect{
\begin{bmatrix} 
\hat{x}^{t|t}\\
\hat{d}^{t|t}
\end{bmatrix}
= 
\begin{bmatrix} 
\hat{x}^{t|t-1}\\
\hat{d}^{t|t-1}
\end{bmatrix}
+
K
\left(
y^t - 
\begin{bmatrix} 
C \\ C_d
\end{bmatrix}^T  
\begin{bmatrix} 
\hat{x}^{t|t-1}\\
\hat{d}^{t|t-1}
\end{bmatrix}  
\right)
}
\end{align}
% \begin{align} \label{eqn:observer_est}
% \vect{
% \underbrace{
% \begin{bmatrix} 
% \hat{x}^{t|t}\\
% \hat{d}^{t|t}
% \end{bmatrix}
% }_{\hat{z}^{t|t}}
% = 
% \underbrace{
% \begin{bmatrix} 
% \hat{x}^{t|t-1}\\
% \hat{d}^{t|t-1}
% \end{bmatrix}
% }_{\hat{z}^{t|t-1}}
% +
% \underbrace{
% \begin{bmatrix} 
% K_x\\
% K_d
% \end{bmatrix}
% }_{K}
% \left(
% y^t - 
% \begin{bmatrix} 
% C & C_d
% \end{bmatrix}  
% \begin{bmatrix} 
% \hat{x}^{t|t-1}\\
% \hat{d}^{t|t-1}
% \end{bmatrix}  
% \right)
% }
% \end{align}
where $\vect{K}$ is the gain matrix for the augmented state and disturbance variable. For simplicity, we adopt a fixed gain structure for the gain matrix $\vect{K}$. 
% We adopt a rolling window least-square approach in determining these gains where the gain matrix $\vect{K}$ is the minimizer of $\Vert \vect{y^t - K\hat{y}^t }\Vert_F$.
% These gains can be determined using a rolling window least square approach as the matrix that minimizes the 

This observer integrated MIPC model in \eqref{eqn:swing_ss_dis_aug} and \eqref{eqn:observer_est} replaces the linear model in \eqref{eqn:swing_ss_red}, with the augmented state used in place of the original states $\vect{x^t}$ and the rest of the algorithm follows through for the constrained and unconstrained case. 

\subsection{Limited Communication}
% In modern power systems, wide area measurement system (WAMS) data, consisting of sensors and communication infrastructures, is increasingly available \cite{phadke2018phasor} as most balancing authorities require that new generators, greater than a certain MW, have phasor measurement units (PMU) installed \cite{pjm}. There isn't still a guarantee that all generators have such information or when there is a communication failure.
While wide area measurement and communication systems are becoming increasingly available \cite{phadke2018phasor}, many networks still do not yet have real-time communication capabilities.  Even for system with these types of infrastructure, there is always the possibility of communication issues.  

To tackle the issue of limited communication, we assume that the initial state measurements of the generators is available. For example, these can be conveyed using the existing SCADA system every two to four seconds. The augmented state and disturbance estimate in \eqref{eqn:observer_est} is then used in estimating the full state and disturbance. The only difference is in the gain used since the structure and dimension of the gain $\vect{K}$ will change depending on the number of generators with available state information, that is, the dimension of $\vect{y^t}$. % As with the state and disturbance estimation case, these gains are also selected as fixed gains. More sophisticated gain structures will be explored in future works.
% As with the state and disturbance estimation case, these gains are also selected using the least square approach. More sophisticated gain structures will be explored in future works.
The key idea here is that the mismatch between the evolved initial state of the generators with limited communication and what the state should be if there was communication is reflected as a disturbance in the network and can be estimated using the measurements from available generators. 

% The discrete form can be computed as $A = e^{\hat{A}Ts} \ \text{and}  \ B = \hat{A}^{-1}e^{(\hat{A}Ts - I)} \hat{B}$

% According to the observer stability theorem in \cite{pannocchia2003disturbance}, \eqref{eqn:swing_ss_dis_aug} is observable if and only if $(\vect{C, \bar{A}})$ in \eqref{eqn:observer_plant} is observable and 
% $ \vect{\begin{bmatrix}
% \bar{A} - I & B_h \\
% C & C_h   
% \end{bmatrix}} $ in \eqref{eqn:swing_ss_dis_aug} is full column rank. Therefore $\vect{B_h}$ and $\vect{C_h}$ should be chosen to such that this observability condition holds.

\section{Case Studies} \label{section:results}
In this section, we validate the performance and versatility of the MIPC controller by testing it on the popular IEEE New England 10-machine 39-bus (NE39) system used for power systems dynamics stability studies~\cite{ramos2015benchmark}. We study scenarios including constraints on the power and energy output of the IBR, noisy measurements and limited communication. Under each scenario, a large disturbance in the form of a partial generating capacity loss is applied to a generator in the network to initiate an event that can lead to a marked frequency decline. The performance metrics for the controller is its ability to maintain the frequency deviation within a small range, quickly recovering to the nominal frequency value and limiting the ROCOF.

\begin{figure}[htbp]
\centering
\includegraphics[scale = 0.8]{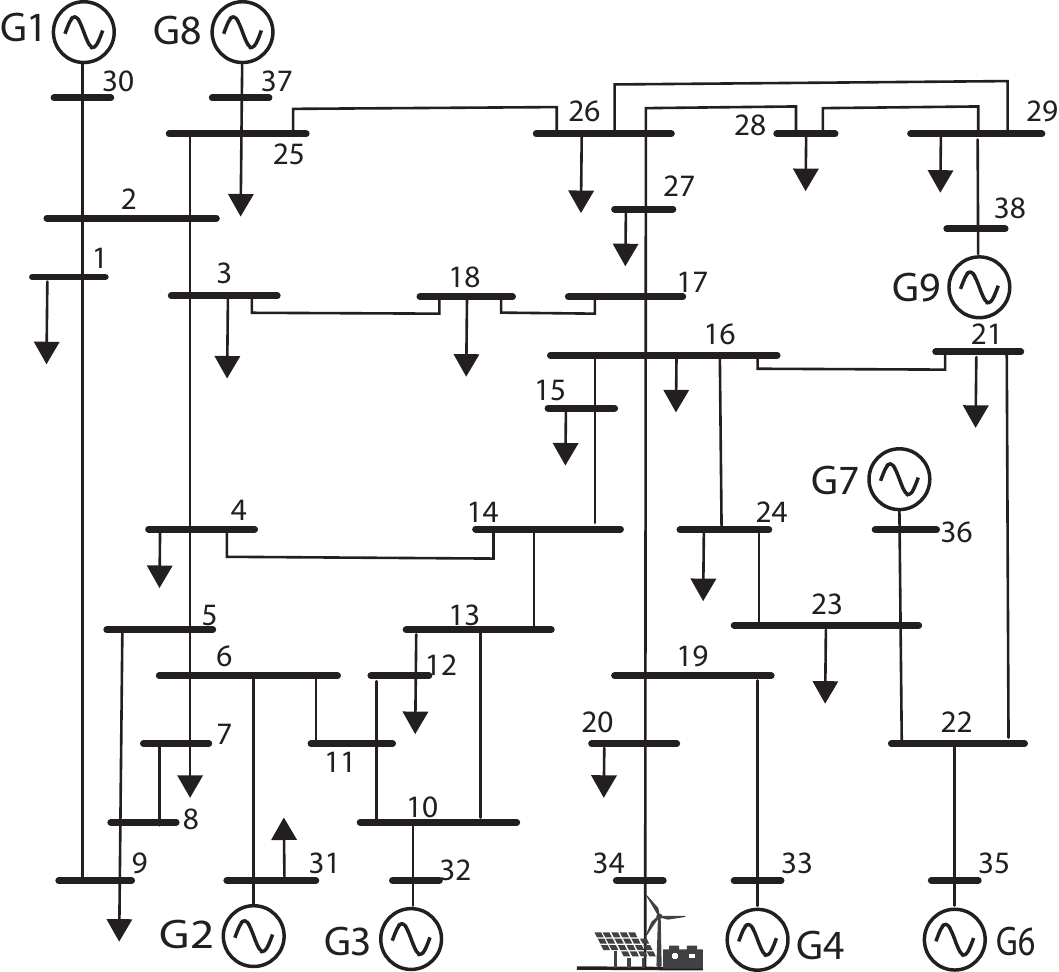}
\caption{New England 39-bus system schematic with  generator 5 at bus 34 replaced with an IBR of equal aggregated capacity. }
\label{fig:ne39_network}
\end{figure}

The network is transformed into a low-inertia network by removing the interconnection to the rest of the US network and replacing generator 5 at bus 34 with an IBR whose total aggregated capacity equal to the replaced generator as shown in Fig. \ref{fig:ne39_network}, and reduced to an equivalent network using Kron reduction. The integrated IBR can be made up of either solar or wind but will be coupled with energy storage devices to guarantee the availability of the required capacity for frequency control. The step size for the discrete simulation is set to 0.05 (50ms) and the simulation starts of with the network in steady state, after which a disturbance of $60\%$ loss of capacity is applied to the fourth generator (\emph{G4}) located at bus 33 from 0.5 to 5 seconds.

The performance of the proposed controller is compared to that of an optimally tuned VSM controller discussed in Section~\ref{subsection:vsm} with computed optimal gain coefficients of $K_\text{m} = 201.7$ and $K_\text{d} = 520$ according to \cite{AdemolaIdowu2018OptimalDO}. Note that the VSM controllers  simply saturates when it hits its power or energy limits.  

\subsubsection{Unconstrained Scenario}
\begin{figure}[htbp]
    \centering
    \includegraphics[width = 1\columnwidth, center]{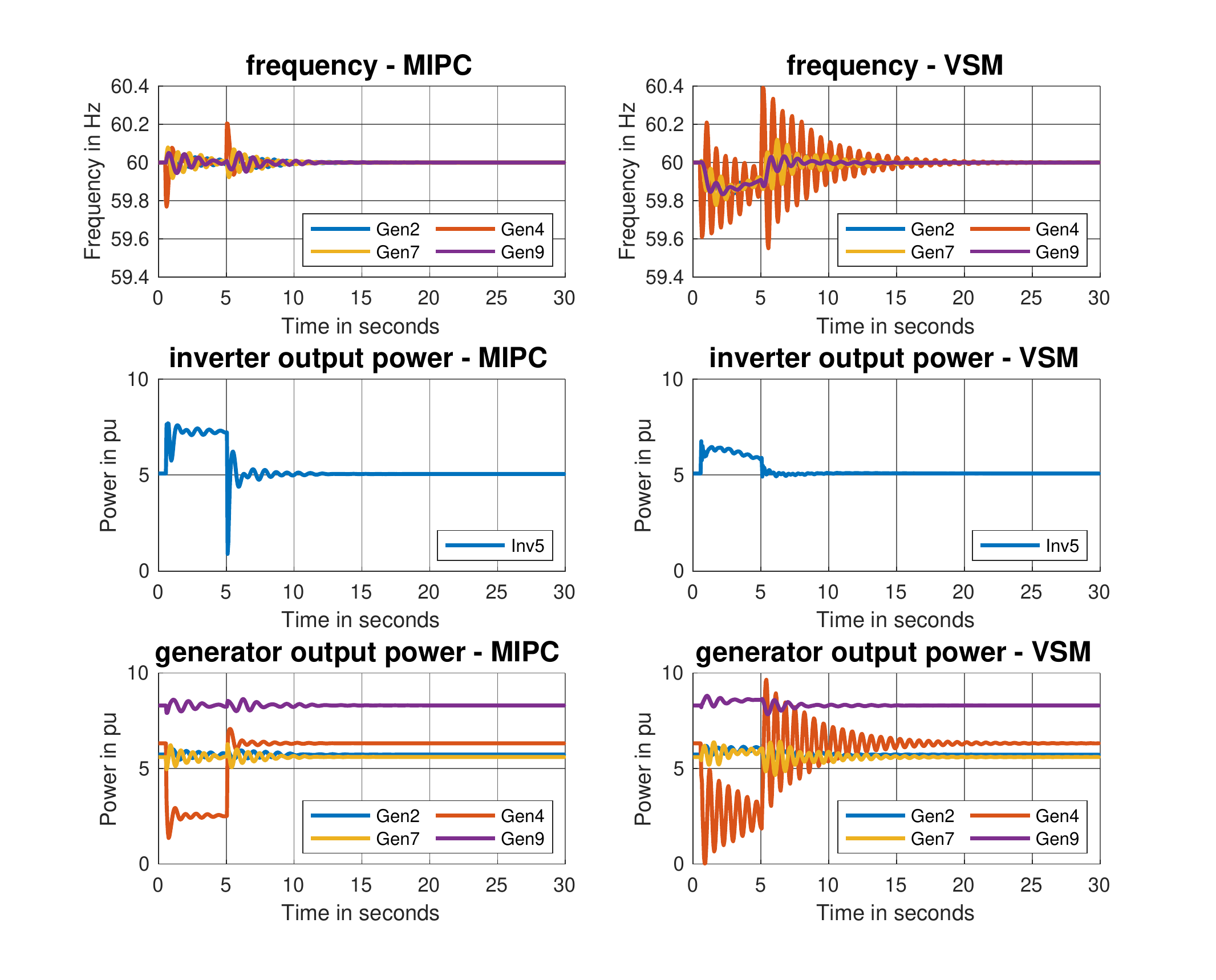}
    \caption{Comparison of MIPC and VSM control strategies for an unconstrained scenario in a NE39 network. The plots show the frequency and generator output power for some select generators and the IBR's output power. The MIPC outperforms the VSM in keeping the frequencies within limits.}
    \label{fig:ne39_uncon}
\end{figure}

Fig. \ref{fig:ne39_uncon} shows the generator frequencies and IBR output power of the MIPC and VSM where there are not energy nor power limits. For a clearer viewing, only the frequency response of the second (slack), fourth (disturbed), seventh and ninth generator are shown. The proposed MIPC controller keeps the frequencies within 0.2 Hz of nominal and restores the frequency to its nominal value, while under the VSM controller, the frequency varies by about 0.4 Hz and oscillates for a much longer time. 
% longer due to the inability of the VSM to anticipate the impact of the inter-area swing, that is a groups of generators swinging against each other \cite{shim2017synchronization}. 
% width=1.2\textwidth,center

\begin{figure}[htbp]
    \centering
    \includegraphics[width = 1\columnwidth, height=3.5cm, center]{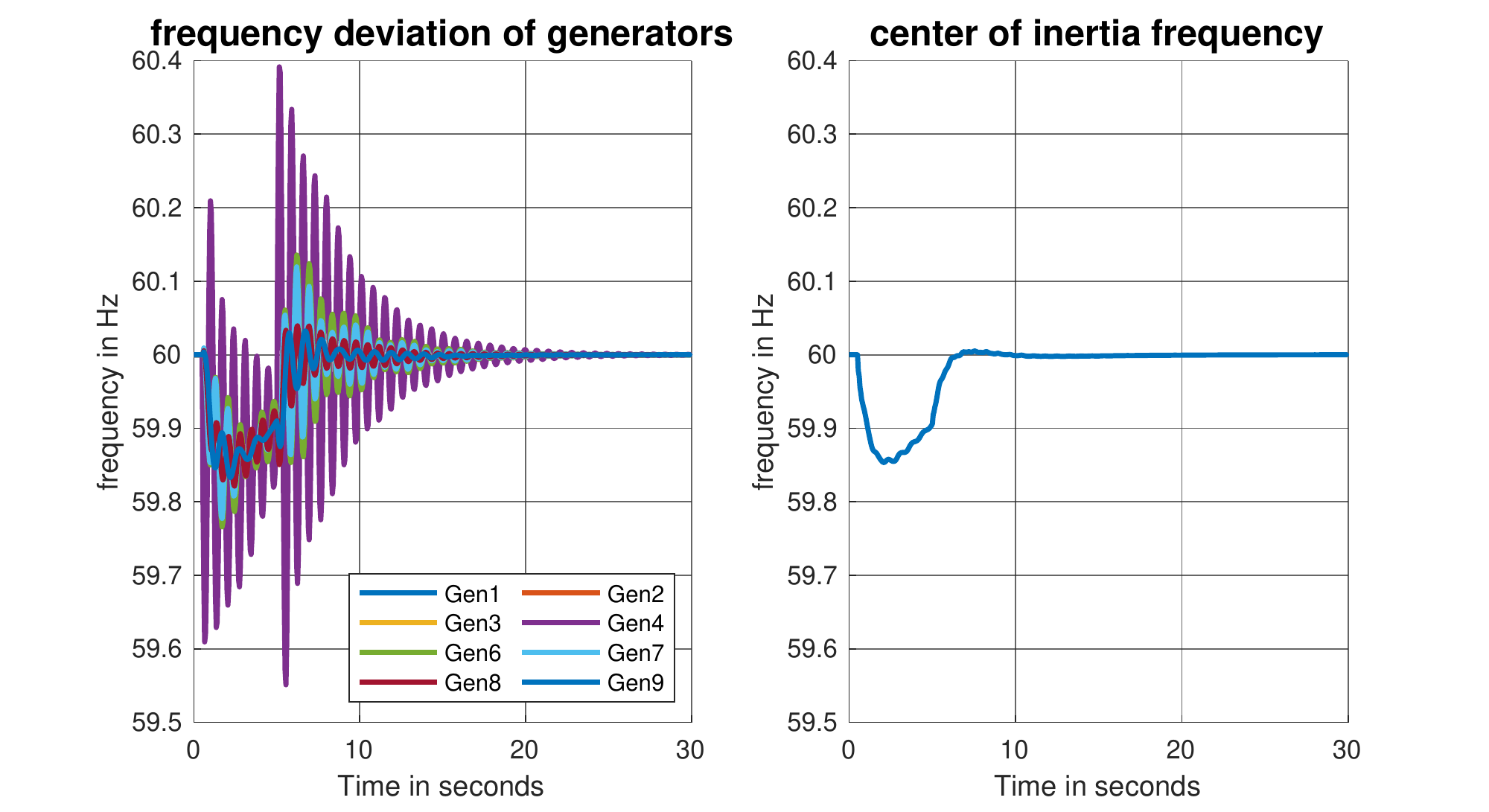}
    \caption{Frequencies of the individual generators (left) and the frequency of the center of inertia (right).}
    \label{fig:speed_compare}
\end{figure}

This shows the look-ahead and adaptive nature of the proposed MIPC controller enables it outperform the VSM by utilizing the resources available to satisfy the given objective. It overcomes a particular issue of when the local frequency measurement becomes misleading a network with different coherent areas swinging against each other. Fig. \ref{fig:speed_compare} shows the center of inertia frequency given by \eqref{eqn:local_freq} when utilizing the VSM control structure compared to the true frequency of each generators, which the former is misleadingly small. 

% It's interesting to note that Fig.~\ref{fig:ne39_uncon} also implies that the performance of an IBR is strictly better than that of a synchronous generator, since the VSM mimics the response of a synchronous generator with optimized inertia and droop coefficients. Therefore, replacing conventional generators by renewable resource does not necessarily mean the frequency response is worse. Rather, if the resource can be optimized, then much better responses are possible.

%\subsubsection{Power and Energy Constraints}
%Figs. \ref{fig:ne39_power} 
%% and \ref{fig:ne39_energy} 
%shows select generators' frequency and output power, and also the IBR output power for a power MIPC and VSM, respectively. The maximum power limits at each time step was set to $7$pu.
%% while the total energy limit was set to $70$pu.
%For a clearer viewing, only the frequency response of the second (slack), fourth (disturbed), seventh and ninth generator are shown. Compared to the VSM, the MIPC is still able to limit the frequency deviation to about 0.3 Hz while keeping the system frequency from rapid oscillation as seen in the VSM case. 
%% for the limited power case and 0.4Hz for the limited energy case 
%This expected since the MIPC is able to integrate the resource constraints into its optimization but is very hard to achieved using a VSM controller since it lacks an explicit optimization step to deal with hard constraints. Similar performances are observed in the energy-limited setting, which we do not show because of space constraints. 
%% therefore unable to utilize the full resource at its disposal.

\subsubsection{Power and Energy Constraints}
\begin{figure}[htbp]
    \centering
    \includegraphics[width = 1\columnwidth, center]{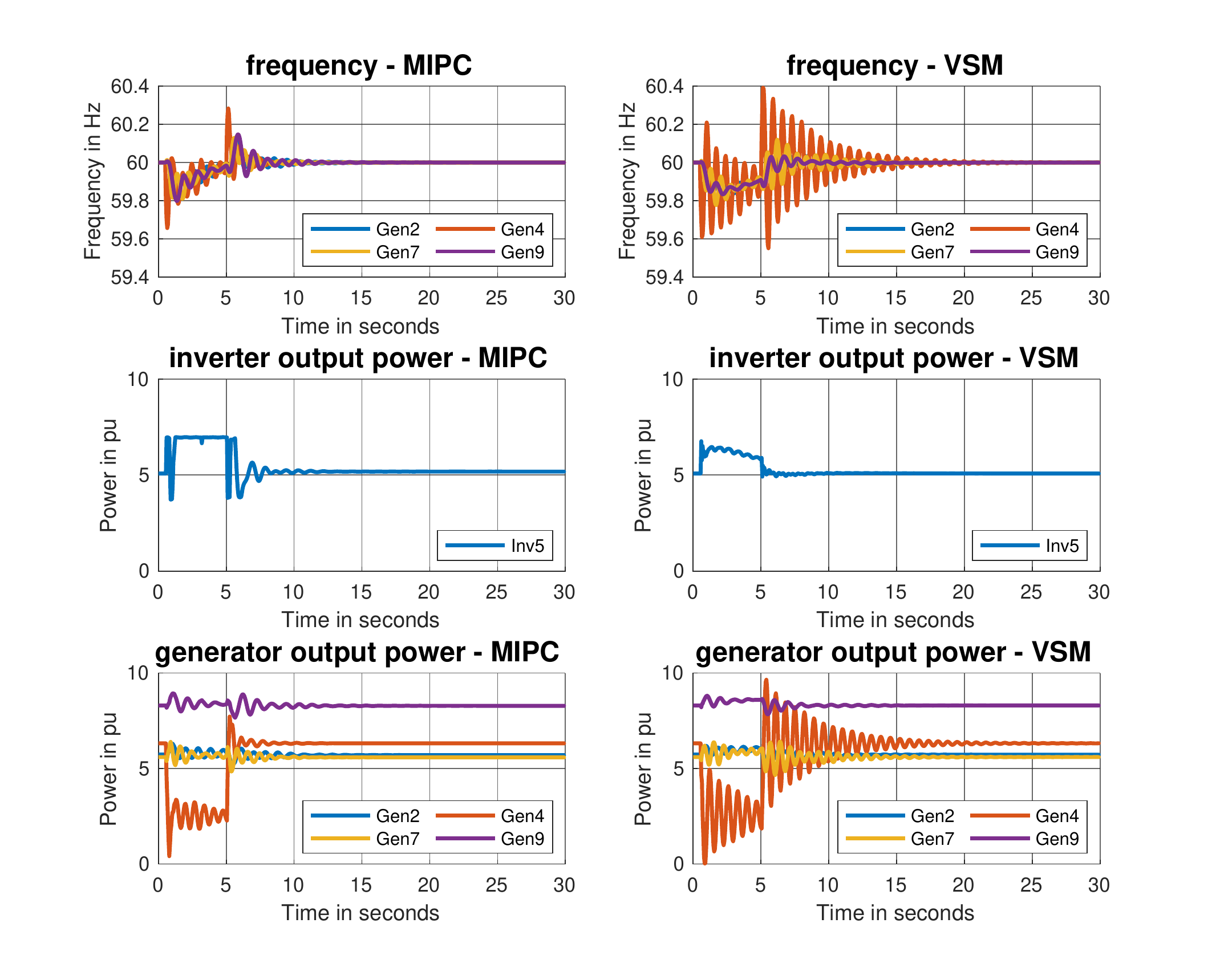}
    \caption{Comparison of MIPC and VSM control strategies for a power constrained scenario in a NE39 network. The MIPC controller is able to adaptively change its power output to minimize frequency deviations while respecting the power limits}
    \label{fig:ne39_power}
\end{figure}

\begin{figure}[htbp]
     \centering
     \includegraphics[width = 1\columnwidth, center]{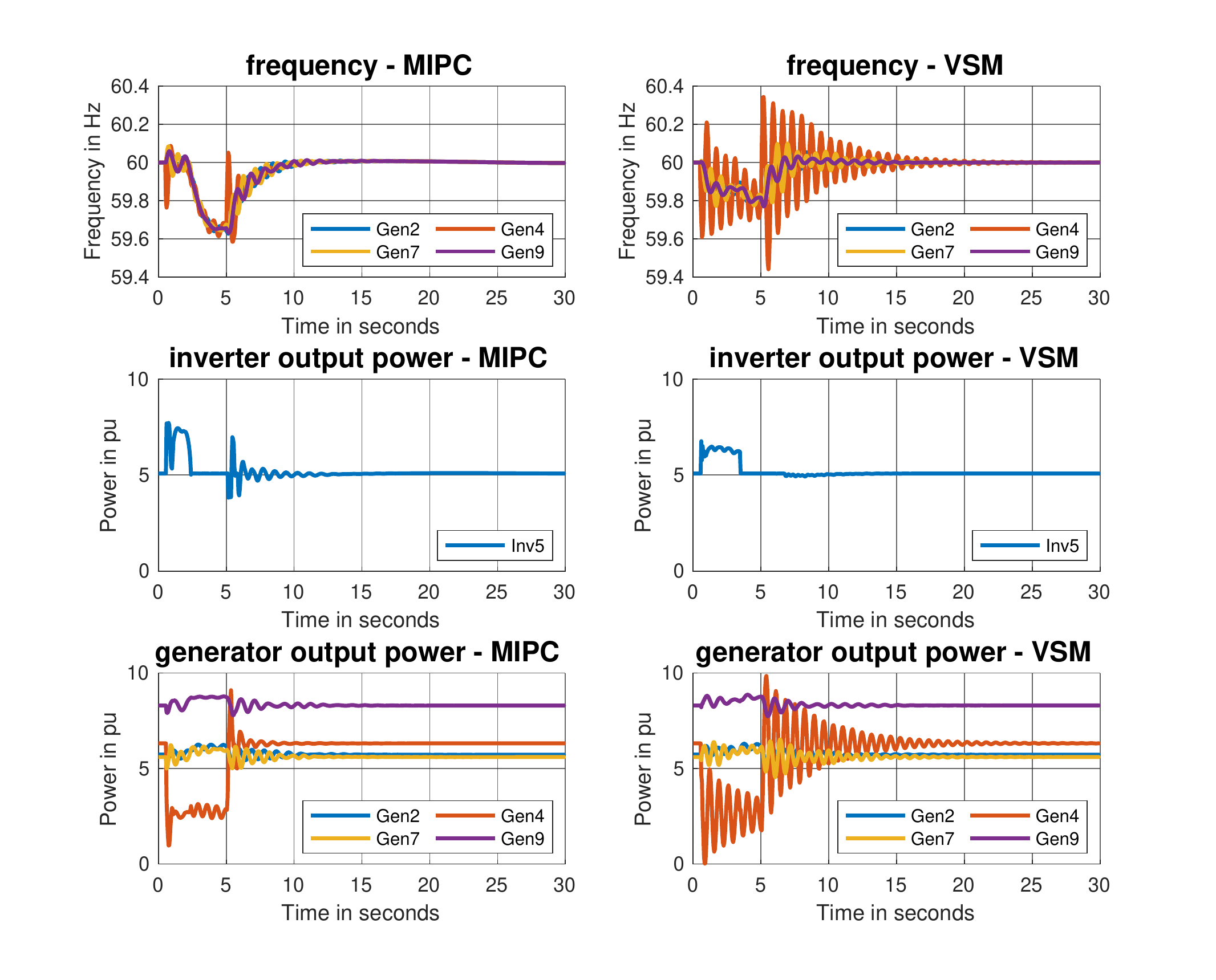}
     \caption{Comparison of MIPC and VSM control strategies for an energy constrained scenario in a NE39 network. The MIPC controller is also able to adaptively change its power output to ensure that the energy limits over its operation horizon are respected.}
     \label{fig:ne39_energy}
 \end{figure}

Figs. \ref{fig:ne39_power} and \ref{fig:ne39_energy} shows select generators' frequency and output power, and also the IBR output power for a power MIPC and VSM, respectively. The maximum power limits at each time step was set to $7$pu while the total energy limit was set to $70$pu. For a clearer viewing, only the frequency response of the second (slack), fourth (disturbed), seventh and ninth generator are shown. Compared to the VSM, the MIPC is still able to limit the frequency deviation to about 0.3 Hz for the limited power case and 0.4Hz for the limited energy case, while keeping the system frequency from rapid oscillation as seen in the VSM case. 
This expected since the MIPC is able to integrate the resource constraints into its optimization but is very hard to achieved using a VSM controller since it lacks an explicit optimization step to deal with hard constraints. Similar performances are observed in the energy-limited setting, which we do not show because of space constraints. 
% therefore unable to utilize the full resource at its disposal.

\subsubsection{Robustness of the Controller}
\begin{figure}[htbp]
    \centering
    \includegraphics[width = 1\columnwidth, center]{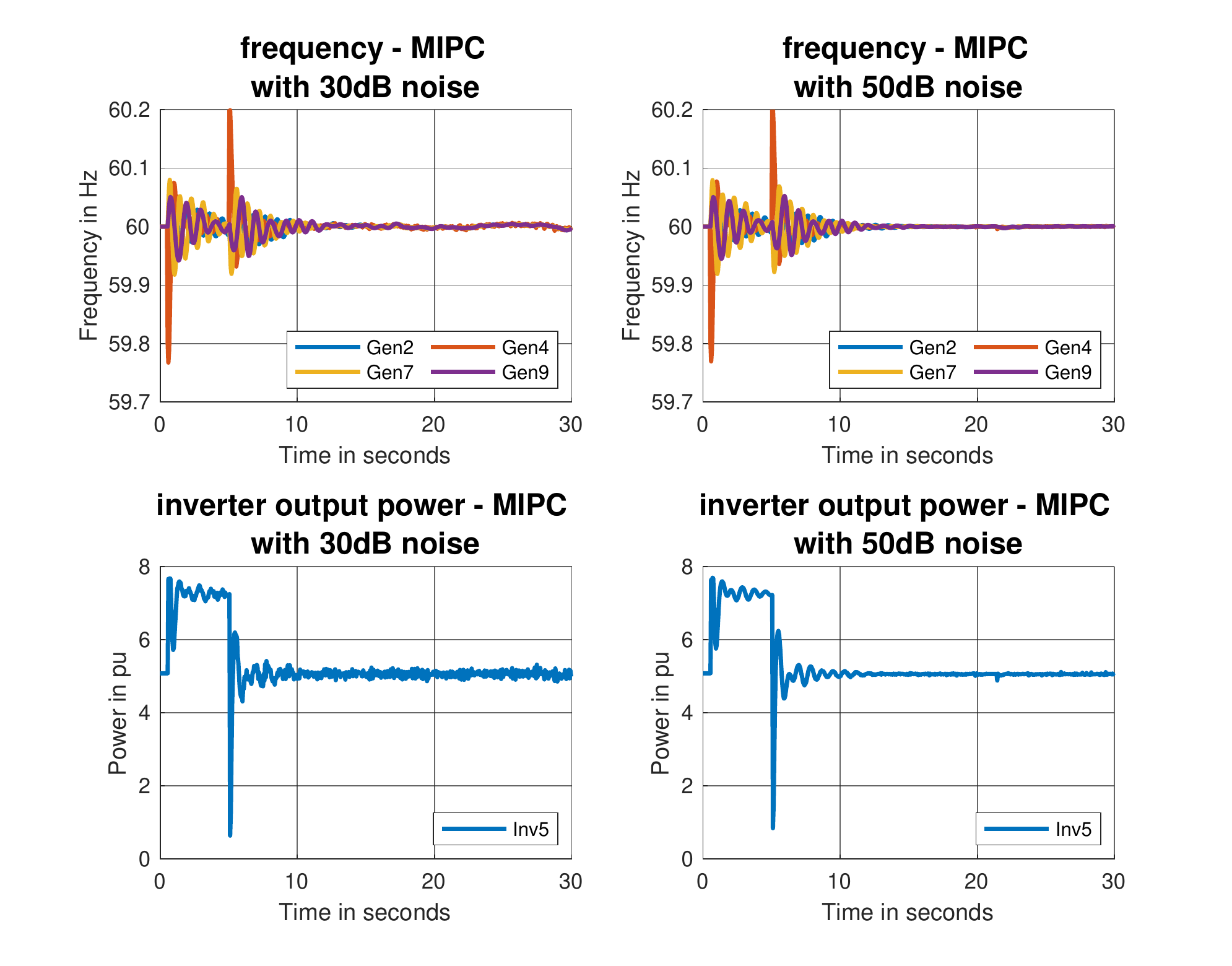}
    \caption{Comparison of robust MIPC under SNR of 30dB and 50dB. The MIPC controller is still able achieve a smooth response while keeping the frequency within limit even in a worst-case noise scenario.}
    \label{fig:ne39_noise}
\end{figure}

Fig.~\ref{fig:ne39_noise} demonstrates the robustness of the MIPC controller to noise, model mismatch and external disturbances to the system with the incorporation of the observer model in \eqref{eqn:observer_est}. 
% According to PMU standards in \cite{8577045}, the total vector error of a PMU measurement should be $< 1\%$($\sim 40$dB signal-to-noise ration (SNR)) while~\cite{ghiga2017phasor} suggested that the SNR of PMU measurements can vary between 30 to 65 dB.
We model noisy measurements by adding Gaussian noise to create SNRs of 30dB and 50dB, respectively representing the worst-case and an average-case SNR scenarios~\cite{8577045,ghiga2017phasor}. Fig.~\ref{fig:ne39_noise} shows that noise has very little impact to the performance of the MIPC (even under 30 dB of SNR). Of course, the observer plays an important role in this robustness to noisy measurements.

\subsubsection{Limited Communication}
\begin{figure}[htbp]
    \centering
    \includegraphics[width = 0.9\columnwidth, center]{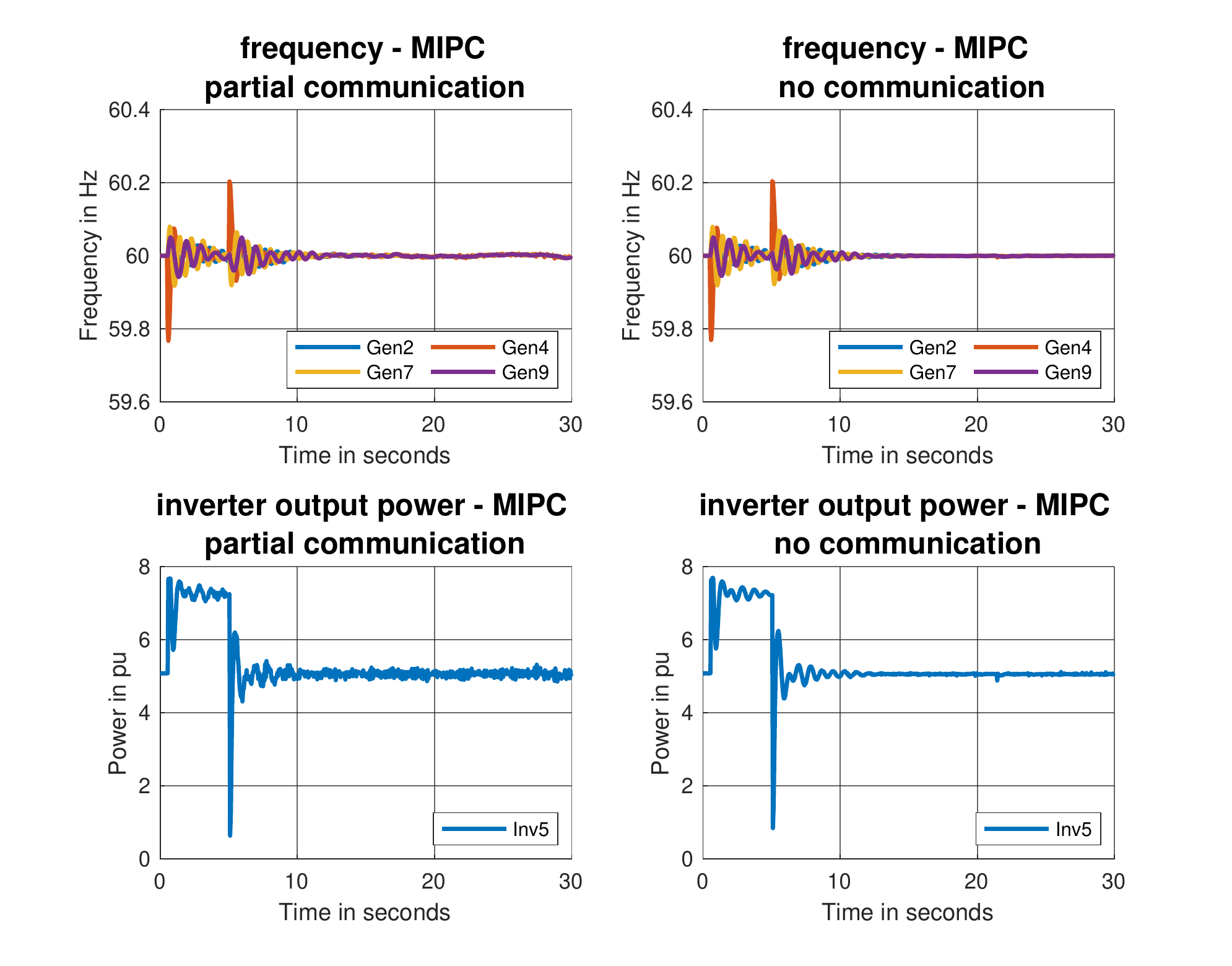}
    \caption{MIPC control actions under a partial and no-communication scenario.}
    \label{fig:ne39_comm}
\end{figure}

We further test the performance of the MIPC in a limited communication scenario. In the partial communication scenario, we assume that measurements can be received from the generators G3, G4, G6, and G7 while only initial state measurements is received from the generators G1, G2, G8 and G9 in Fig. \ref{fig:ne39_network} (so faulted generator (G4) is able to communicate with the inverter). 
The left-side plot of Fig. \ref{fig:ne39_comm} shows the MIPCs performance under the partial communication scenario. By communicating with some buses, the MIPC is able to reconstruct enough of the system-level information to make the computations at the MIPC useful. 

The right-side plot of Fig. \ref{fig:ne39_comm} shows the no-communication scenario, where none of the generators can exchange information with each other. Here the MIPC performs much like a VSM. This is expected since without communication, the best MIPC can do is to utilize its local measurements to take actions. 
% This is because the observer model in \eqref{eqn:observer_est} is able to estimate the true system state and uses difference in the measured and estimated states of the known communication to make up for the limited communication.

\subsubsection{Delayed Communication}
\begin{figure}[htbp]
    \centering
    \includegraphics[width = 0.9\columnwidth, center]{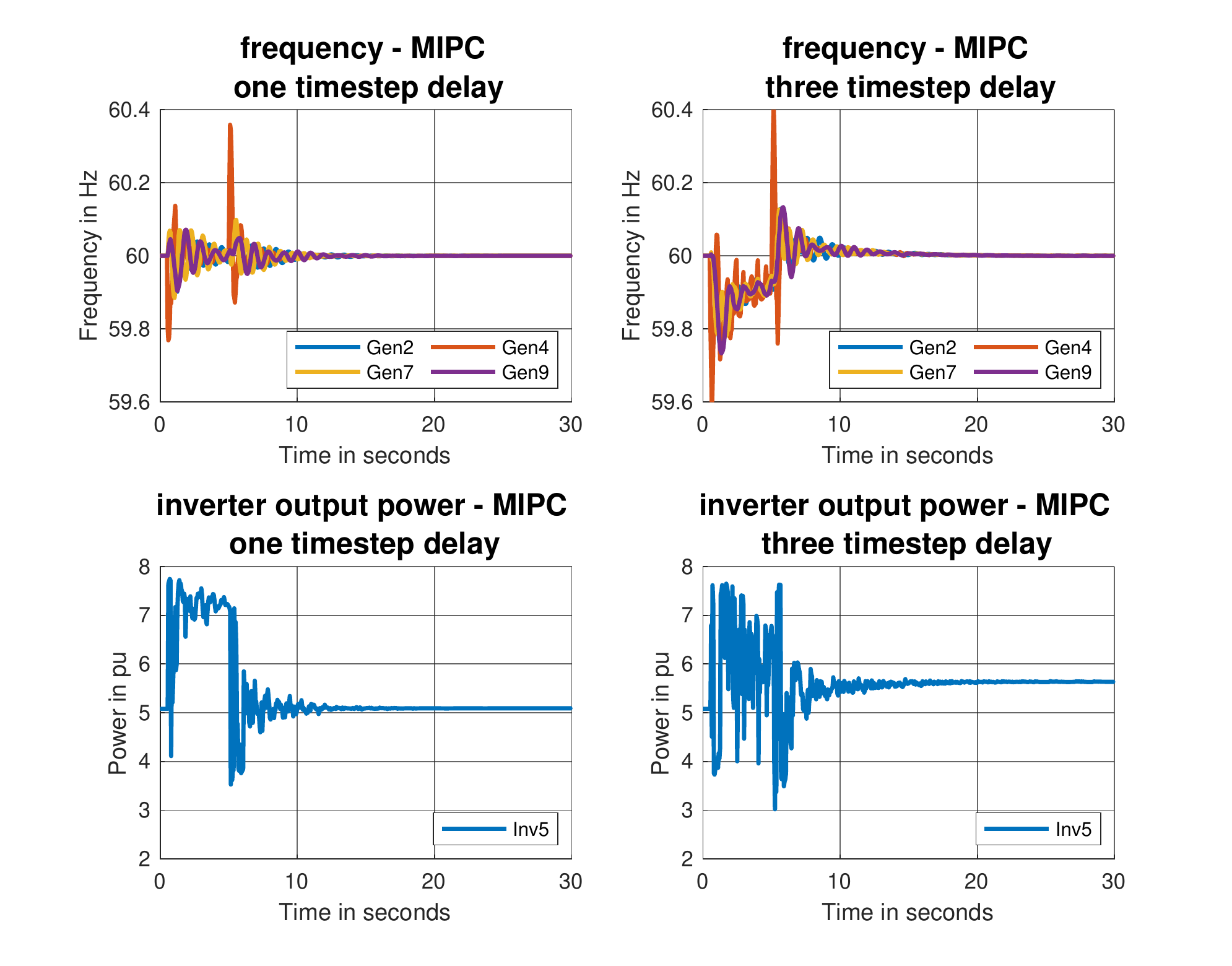}
    \caption{MIPC control actions under a one- and three- time step delayed communication scenario.}
    \label{fig:ne39_delay}
\end{figure}

In the previous scenarios, we assumed that the measurements are received in time for the controller to perform its computations. In this delayed communication scenario, we assume that there is a delay in measurement data received from some generators in the network. Fig.~\ref{fig:ne39_delay} shows the MIPC performance when a delay in measurement received is applied at random to any four generators thus creating a possibility for the controller to receive delayed measurements from the faulted generator (G4). The left-side plot of Fig. \ref{fig:ne39_delay} shows a one time step delay, that is, a delay of 50ms, while the right-side plot shows a three time step delay, that is, a delay of 150ms.  It can be observed that the MIPC controller is still able to determine the appropriate inverter set-point to drive the system to stability albeit with deteriorating performance as the time lag increases.

\section{Conclusion} \label{section:conclusion}
In this paper, we proposed a novel control strategy called the Inverter Power Control that optimally determines the active power set-point for an inverter-based resource in real-time. Using a model predictive control framework, hard power and energy constraints are considered explicitly in the optimization process. We show via simulation on a test system the superiority of the proposed controller in comparison to the optimally tuned virtual synchronous machine, under both noisy and limited communication settings. Our future work explores enhancing the controller to function in a large network with multiple IBRs, integrating model identification techniques and robustness to communication delays.

% Our future works include delayed communication, multiple inverters, improved limited communication, model identification to reduce the reliance on an abundance of data.

% \appendix[Proof of the Zonklar Equations]
% or
% \appendix  % for no appendix heading
% \input{appendix.tex}

% \section{}

\bibliographystyle{IEEEtran}
\bibliography{Inverter_Support}

\appendix  % for no appendix heading
\raggedbottom\sloppy
\section*{Constant Gain Matrix for MIPC} \label{appendix:ipc_gain}
The matrices $\vect{H}$ and $\vect{F}$ in \ref{subsection:unconstrained} can be obtained as follows: writing the linear system model in \eqref{eqn:swing_ss_red} for $N$ time steps ahead in matrix form, we have:
\begin{align} \label{eqn:stacked_dyn}
\resizebox{.95\columnwidth}{!}{$
\underbrace{
\begin{bmatrix} 
\vect{x^0}\\
\vect{x^1}\\
\vect{x^2}\\
\vdots\\
\vect{x^N}
\end{bmatrix}
}_{\vect{x}}
=
\underbrace{
\begin{bmatrix}
\vect{0} & \vect{0} & \hdots & \vect{0}\\
\vect{\bar{B}} & \vect{0} & \hdots & \vect{0}\\
\vect{\bar{A}\bar{B}} & \vect{\bar{B}} & \hdots & \vect{0}\\
\vdots & \vdots & \ddots &\vdots\\
\vect{\bar{A}}^{N-1}\vect{\bar{B}} & \vect{\bar{A}}^{N-2}\vect{\bar{B}} & \hdots & \vect{\bar{B}}\\
\end{bmatrix}
}_{\vect{S}}
\underbrace{
\begin{bmatrix} 
\vect{u^0}\\
\vect{u^1}\\
\vdots\\
\vect{u^{N-1}}
\end{bmatrix}
}_{\vect{u}}
+
\underbrace{
\begin{bmatrix} 
\vect{I}\\
\vect{\bar{A}}\\
\vect{\bar{A}}^2\\
\vdots\\
\vect{\bar{A}}^N
\end{bmatrix}
}_{\vect{M}}
\vect{x^0}
$}
\end{align}
The objective function in \eqref{eqn:opt_inv} can then be written in terms of the state variable as:
\begin{align}
\vect{y^T Q_1 y}
& = 
\vect{(C x)^T Q_1 (C x)}
= \vect{x^T \underbrace{C^T Q_1 C}_{\hat{Q}_1} x} \\ \nonumber
& =
\begin{bmatrix} 
\vect{x^0}\\
\vect{x^1}\\
\vect{x^2}\\
\vdots\\
\vect{x^N}
\end{bmatrix}
\vect{
\underbrace{
\begin{bmatrix}
\hat{Q}_1 & 0 & 0 & \hdots &0\\
0 & \hat{Q}_1 & 0 & \hdots &0\\
0 & 0 & \hat{Q}_1 & \hdots &0\\
\vdots & \vdots & \ddots &\vdots\\
0 & 0 & 0 & \hdots &\hat{Q}_1\\
\end{bmatrix}
}_{\tilde{Q}_1}
}
\begin{bmatrix} 
\vect{x^0}\\
\vect{x^1}\\
\vect{x^2}\\
\vdots\\
\vect{x^N}
\end{bmatrix} \\
& = \vect{(Su + Mx^0)^T \tilde{Q}_1 (Su + Mx^0)}
\end{align}
% Let 
% \begin{align}
% \vect{\Theta} & = \vect{S_{1:N-1} - S_{2:N}} \\ \nonumber
% \vect{\Gamma} & = \vect{M_{1:N-1} - M_{2:N}} 
% \end{align}
% and
% \begin{align}
% \vect{\bigtriangleup x^0} & = \vect{x^0 - x^1} \\ \nonumber
% \vect{\bigtriangleup x^1} & = \vect{z^1 - x^2} \\ \nonumber
% \vdots \\
% \vect{\bigtriangleup x^{N-1}} & = \vect{x^{N-1} - x^N},
% \end{align}
% Let $\vect{\Theta} = \vect{S[1:N-1;1:N] - S[2:N;1:N]}$, $\vect{\Gamma} = \vect{M[1:N-1] - M[2:N]}$, and $\vect{\bigtriangleup x} = \vect{x[1:N-1] - x[2:N]}$ such that:
Let
\begin{align}
    \vect{\Theta} & = \vect{S[0:N-1;1:N] - S[1:N;1:N]} \\ \nonumber
    \vect{\Gamma} & = \vect{M[0:N-1] - M[1:N]}\\ \nonumber
    \vect{\bigtriangleup x} & = \vect{x[0:N-1] - x[1:N]}
\end{align}
such that
\begin{align}
\vect{\bigtriangleup y^T Q_2 \bigtriangleup y}
& = 
\vect{(C \bigtriangleup x)^T Q_2 (C \bigtriangleup x)} \\ \nonumber
& = \vect{\bigtriangleup x^T \underbrace{C^T Q_2 C}_{\hat{Q}_2} \bigtriangleup x} \\ \nonumber
& = \vect{(\Theta u + \Gamma x^0)^T \tilde{Q}_2 (\Theta u + \Gamma x^0)}
\end{align}
Therefore \eqref{eqn:opt_inv} becomes:
\begin{equation}
    \begin{aligned}
    J & = \frac{1}{2} \Big(\vect{(Su + Mx^0)^T \tilde{Q}_1 (Su + Mx^0)} \\
    & \qquad + \vect{(\Theta u + \Gamma x^0)^T \tilde{Q}_2 (\Theta u + \Gamma x^0)} \Big) \\ 
    & = \frac{1}{2} \vect{{x^0}^T \Big[\underbrace{M^T \tilde{Q}_1 M + \Gamma^T \tilde{Q}_2 \Gamma}_{G}\Big]x^0}\\ 
    & \qquad + \frac{1}{2} \vect{u^T \Big[\underbrace{S^T \tilde{Q}_1 S + \Theta^T \tilde{Q}_2 \Theta}_{H}\Big]u}\\ 
    & \qquad + \vect{{x^0}^T \Big[\underbrace{M^T \tilde{Q}_1 S + \Gamma^T \tilde{Q}_2 \Theta}_{F}\Big]u} \\
    \label{eqn:uncon_obj}
    J & = \frac{1}{2}\vect{{x^0}^T G x^0} + \frac{1}{2}\vect{u^T H u + {x^0}^T F u}  
    \end{aligned}
\end{equation}

For optimality, $\nabla J_u = \vect{H u + F x^0} = 0$, such that the optimal control action at a given start point for $N$ time horizon ahead becomes: 

\begin{equation} 
    \begin{aligned}
        \vect{u^*} = \vect{-H^{-1} F^T x^0}.
    \end{aligned}
\end{equation}

\section*{Power Limit} \label{appendix:power}
The linear constraint equation for the IBR power limit in \eqref{eqn:opt_qp_power} can be derived by stacking \eqref{eqn:lin_ibr_power} for a $N$ step control horizon as follows:
\begin{equation} \label{eqn:ibr_power_stacked}
\begin{aligned}
\underbrace{
\begin{bmatrix}
P_{\text{ibr},k}^1\\
P_{\text{ibr},k}^2\\
\vdots\\
P_{\text{ibr},k}^N\\   
\end{bmatrix}
}_{\vect{P_{\text{ibr},k}}}
& =
\vect{
\underbrace{
\begin{bmatrix}
B_{p1} & 0 & \hdots & 0\\
B_{p1} & B_{p1} & \hdots & 0\\
\vdots & \vdots & \ddots &\vdots\\
B_{p1} & B_{p1} & \hdots & B_{p1}\\
\end{bmatrix}
}_{B_{p1, N}}
\underbrace{
\begin{bmatrix}
\hat{z}^1\\
\hat{z}^2\\
\vdots\\
\hat{z}^N\\   
\end{bmatrix}
}_{\hat{z}}}\\
&+ 
\vect{
\underbrace{
\begin{bmatrix}
B_{p2} & 0 & \hdots & 0\\
B_{p2} & B_{p2} & \hdots & 0\\
\vdots & \vdots & \ddots &\vdots\\
B_{p2} & B_{p2} & \hdots & B_{p2}\\
\end{bmatrix}
}_{B_{p2, N}}
\underbrace{
\begin{bmatrix}
u^0\\
u^1\\
\vdots\\
u^{N-1}\\   
\end{bmatrix}
}_{u}.}
\end{aligned}
\end{equation}

From \eqref{eqn:stacked_dyn}, we have that $\vect{\hat{z} = Su + M\hat{z}^0}$, therefore \eqref{eqn:ibr_power_stacked} becomes,
\begin{equation} \label{eqn:ibr_power_concise}
\begin{aligned}
\vect{P_{\text{ibr},k}}
& =
\vect{B_{p1, N}Su + B_{p1, N}M\hat{z}^0 + B_{p2, N}u} \\
& = \vect{(\underbrace{B_{p1, N}S + B_{p2, N}}_{\tilde{B}_{p1}})u +  \underbrace{B_{p1, N}M}_{\tilde{B}_{p2}} \hat{z}^0.}
\end{aligned}
\end{equation}
The lower power limit can then be written as 
\begin{equation} 
\begin{aligned}
    \vect{ - \tilde{B}_{p1} u \leq  -\tilde{P}_{\text{ibr, min}} + \tilde{B}_{p2} \hat{z}^0} 
\end{aligned}
\end{equation}

and the upper power limit as
\begin{equation} 
\begin{aligned}
    \vect{ \tilde{B}_{p1} u \leq  \tilde{P}_{\text{ibr, min}} - \tilde{B}_{p2} \hat{z}^0}
\end{aligned}
\end{equation}

resulting in a combined form of:
\begin{equation} 
\begin{aligned} 
\vect{
\underbrace{
\begin{bmatrix} 
-\tilde{B}_{\text{p2}} \\
\tilde{B}_{\text{p2}} 
\end{bmatrix}
}_{L}
\vect{u} 
\leq
\underbrace{
\begin{bmatrix} 
-\tilde{P}_{\text{ibr, min}}\\
\tilde{P}_{\text{ibr, max}}
\end{bmatrix}
}_{W}
+
\underbrace{
\begin{bmatrix} 
\tilde{B}_{\text{p1}}\\
-\tilde{B}_{\text{p1}} 
\end{bmatrix}
}_{V} \hat{z}^0.
}
\end{aligned}
\end{equation}
where $\vect{\tilde{P}_{\text{ibr, min}}}$ and $\vect{\tilde{P}_{\text{ibr, max}}}$ are matrices of 
$\vect{\tilde{P}_{\text{ibr, min}}^t}$ and $\vect{\tilde{P}_{\text{ibr, max}}^t}$ stacked together for $N$ horizon.

\section*{Total Energy Limit} \label{appendix:energy}
The linear constraint equation for the IBR total energy limit in \eqref{eqn:opt_qp_energy} can be derived from the power limit in \eqref{eqn:ibr_power_concise} by taking the rolling sum such that:
\begin{equation} 
\begin{aligned}
 \vect{[\tilde{B}_{\text{p1}}u + \tilde{B}_{\text{p2}} \hat{z}^0]^T 1_N \leq \tilde{E}_{\text{ibr, tot}}}   
\end{aligned}
\end{equation}
resulting in
\begin{equation} 
\begin{aligned}
    \vect{\underbrace{1_N^T \tilde{B}_{\text{p1}}}_{\tilde{B}_{\text{e1}}} u \leq \tilde{E}_{\text{ibr, tot}} - \underbrace{1_N^T \tilde{B}_{\text{p2}}}_{\tilde{B}_{\text{e2}}} \hat{z}^0.}
\end{aligned}
\end{equation}
The rate limit, represented as $\epsilon$ in this work, can be incorporated by taking the difference between time steps of the output power in \eqref{eqn:ibr_power_concise} such that:
\begin{equation} 
\begin{aligned}
    \vect{[\tilde{B}_{\text{p1}}u + \tilde{B}_{\text{p2}} \hat{z}^0]_{[1:N-1]} - 
    [\tilde{B}_{\text{p1}}u + \tilde{B}_{\text{p2}} \hat{z}^0]_{[2:N]}} 
\end{aligned}
\end{equation}
resulting in
\begin{equation} 
\begin{aligned}
    \vect{\tilde{B}_{\text{p1}}u + \tilde{B}_{\text{p2}} \hat{z}^0]_{[1:N-1]} - 
    [\tilde{B}_{\text{p1}}u + \tilde{B}_{\text{p2}} \hat{z}^0]_{[2:N]}} 
\end{aligned}
\end{equation}
The combined form of these two constraints is:
\begin{equation} 
\begin{aligned} 
\vect{
\underbrace{
\begin{bmatrix} 
\tilde{B}_{\text{e2}}\\
\tilde{B}_{\text{r2}}
\end{bmatrix}
}_{L}
\vect{u} 
\leq
\underbrace{
\begin{bmatrix} 
\tilde{E}_{\text{ibr, tot}}\\
\epsilon
\end{bmatrix}
}_{W}
+
\underbrace{
\begin{bmatrix} 
- \tilde{B}_{\text{e1}} \\
- \tilde{B}_{\text{r1}}
\end{bmatrix}
}_{V} \hat{z}^0.
}
\end{aligned}
\end{equation}

\end{document}